\documentclass[journal=jpclcd,manuscript=letter,hyperref]{achemso}

\usepackage{chemformula} 
\usepackage[T1]{fontenc} 
\hypersetup{colorlinks=true,allcolors=blue}

\author{Juno Nam}
\author{YounJoon Jung}
\affiliation[Seoul National University]
{Department of Chemistry, Seoul National University, Seoul 08826, South Korea}
\email{yjjung@snu.ac.kr}

\title
  {Enhanced Sampling for Free Energy Profiles with\
   Post-Transition-State Bifurcations}

\begin{document}

\begin{tocentry}
\includegraphics[width=2in]{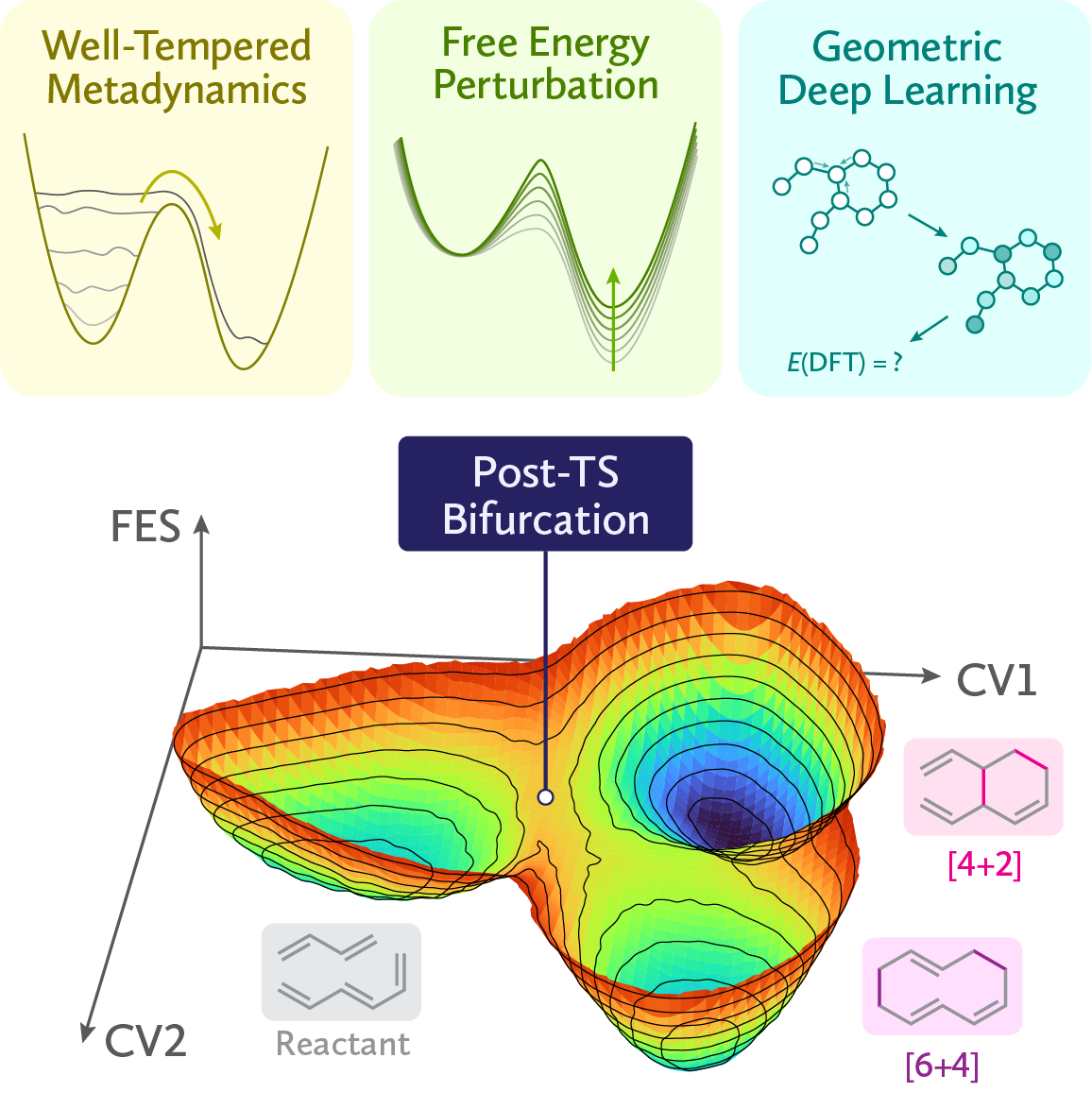}
\end{tocentry}

\begin{abstract}
We present a method to explore the free energy landscapes of chemical reactions with post-transition-state bifurcations using an enhanced sampling method based on well-tempered metadynamics.
Obviating the need for computationally expensive DFT-level ab initio molecular dynamics simulations, we obtain accurate energetics by utilizing a free energy perturbation scheme and deep learning estimator for the single-point energies of substrate configurations.
Using a pair of easily interpretable collective variables, we present a quantitative free energy surface that is compatible with harmonic transition state theory calculations and in which the bifurcations are clearly visible.
We demonstrate our approach with the example of the SpnF-catalyzed Diels--Alder reaction, a cycloaddition reaction in which post-transition-state bifurcation leads to the [4+2] as well as the [6+4] cycloadduct.
We obtain the free energy landscapes for different stereochemical reaction pathways and characterize the mechanistic continuum between relevant reaction channels without explicitly searching for the pertinent transition state structures.  
\end{abstract}



Understanding the selectivity of a reaction is one of the most important challenges in computational chemistry.
Reaction selectivity under kinetically controlled conditions can be predicted by estimating the relative heights of barriers or the free energies of transition states (TSs).
Although this type of prediction is frequently utilized, its use implicitly assumes that competing reactive channels are well separated on the potential energy surface (PES), which is not always the case.
Post-transition-state bifurcations (PTSBs) in chemical reactions occur when reactive trajectories passing through a single ambimodal TS bifurcate into two (or more) different products.\cite{ess2008,rehbein2011,hare2017a}
Such phenomena have been observed for various organic, bioorganic, and organometallic systems.\cite{yamataka1999,caramella2002,ussing2006,ccelebi2007,thomas2008,hong2009,siebert2011,zhang2015,patel2016,hare2017b,hare2017c,yu2017,hare2018,xue2019,liu2020,zhang2021}

The occurrence of bifurcations on PESs means that intrinsic reaction coordinate (IRC) analysis,\cite{fukui1970,fukui1981} which is commonly used to study reactions, is not sufficient to characterize reactive systems featuring PTSBs.
Accordingly, molecular dynamics (MD) simulations---in particular, quasiclassical direct dynamics and quantum mechanics/molecular mechanics (QM/MM) simulations---have been utilized to understand reaction mechanisms and selectivity associated with PTSBs.\cite{zheng2017,chen2018,yang2018a,yang2018b,yang2018c,yang2019,zou2021,jamieson2021,wang2021,qin2021,ito2022}
Although quasiclassical direct dynamics simulations account for product distributions and offer real-time trajectories from which mechanistic properties (e.g., synchronicity) can be probed, the results of such analyses are dependent on the starting TS structures.
It has been reported that PESs in the vicinity of ambimodal transition states may possess a large number of saddle points because of the complexity of the chemical interactions and conformational flexibility of the reactants.\cite{medvedev2017}
In this respect, consideration of the transition state ``ensemble'' is indispensable for obtaining a comprehensive understanding of complicated chemical reactions with PTSBs.

In this study, we aimed to explore free energy profiles in systems characterized by PTSBs by implicitly considering the relevant TS structures using an enhanced sampling method, well-tempered metadynamics (WTMetaD),\cite{barducci2008} with ab initio molecular dynamics (AIMD).\cite{marx2009}
Enhanced sampling methods based on metadynamics\cite{laio2002} have been increasingly utilized in recent years to explore the free energy landscapes of chemical reactions.\cite{piccini2018,rizzi2019,schilling2020,fu2021,trizio2021,raucci2022}
These methods allow efficient exploration of substrate configurations over the reaction barrier on the timescale of AIMD by accumulating a bias potential along predefined collective variables (CVs).
Because of the high computational cost of AIMD, chemical reactions are often simulated at a semiempirical level, and hence the free energy surfaces (FESs) obtained from these dynamics simulations have intrinsically limited accuracy.
Recently, the free energy perturbation (FEP) method has been applied to such simulations\cite{li2018,piccini2019} to obtain more accurate FESs using single-point calculations for sampled configurations at a higher level of theory.
Despite these improvements, as the number of atoms in the reactant increases, the increase in the computational cost of performing single-point calculations for a sufficiently large number of configurations to obtain a FES can be prohibitive.
Motivated by the successful application of geometric deep learning methods for the prediction of various molecular properties,\cite{atz2021} we constructed a FES by training a graph neural network on the DFT calculation results for a small number of configurations sampled by WTMetaD simulation and predicting the energy for a large number of configurations.

\begin{figure}
  \includegraphics[width=0.7\linewidth]{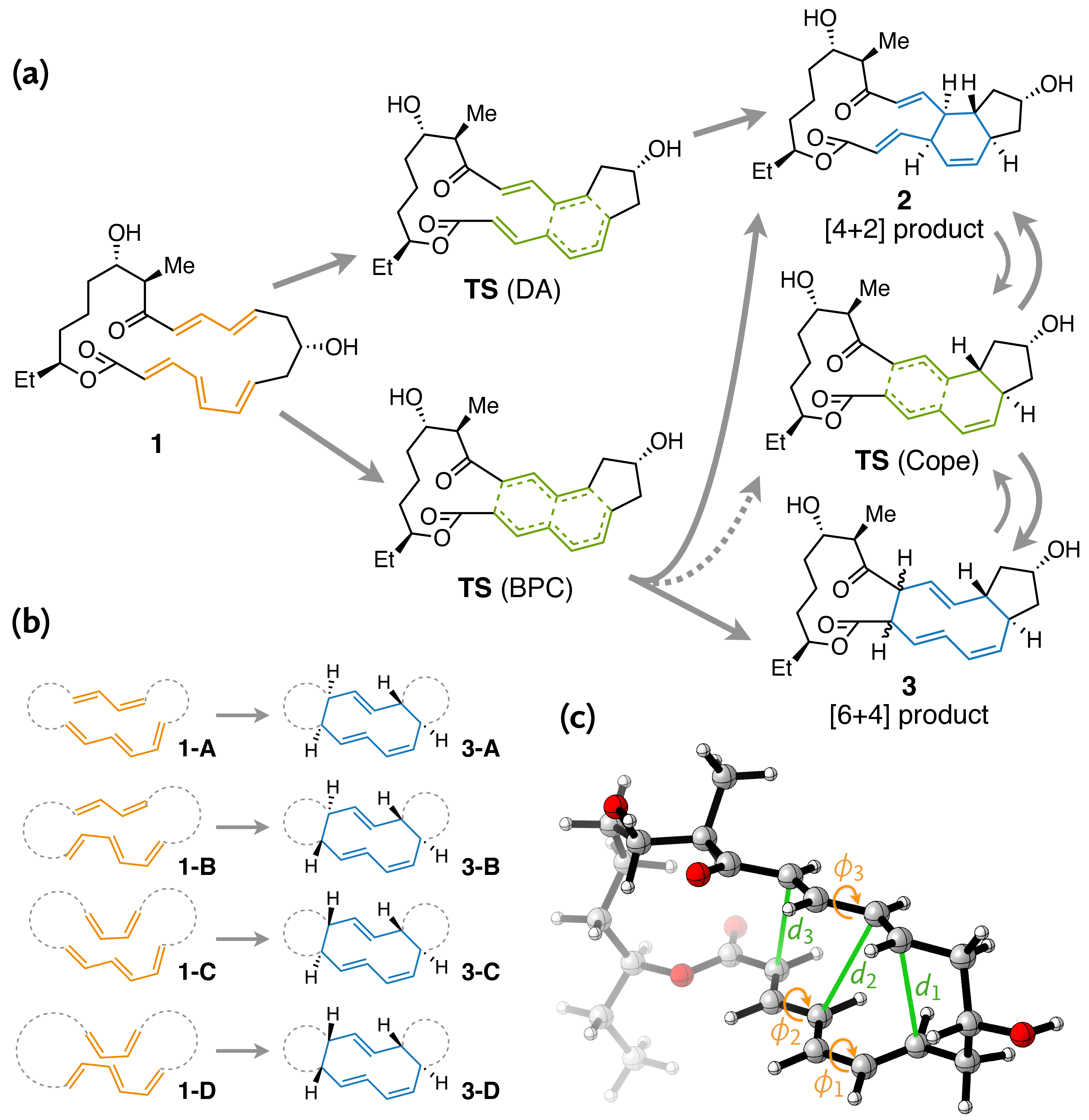}
  \caption{SpnF-catalyzed Diels--Alder reaction.
    (a) Possible cycloaddition routes for the reaction.
    (b) Reactant conformations and [6+4] products with different stereochemistry.
    (c) Relevant geometric parameters (distances and torsional angles) for the substrate.}
    \label{fig:spnf}
\end{figure}

To illustrate the application of our method to explore free energy landscapes with PTSBs, we characterized the ensemble of ambimodal transition states and relevant free energy landscapes of the SpnF-catalyzed Diels--Alder (DA) reaction (Figure~\ref{fig:spnf}a).
The reaction corresponds to a [4+2] cycloaddition step in the biosynthesis of Spinosyn A and is characterized by PTSB because of the existence of ambimodal [6+4]/[4+2] bis-pericyclic (BPC) transition states.\cite{kim2011,fage2015,patel2016}
While several previous studies have focused on the effect of the enzyme and solvent on the reaction pathway and product selectivity (i.e., whether the reaction proceeds via BPC-type or DA-type TSs),\cite{zheng2017,chen2018,yang2018a} Medvedev et al.\cite{medvedev2017} found that more than 300 BPC/DA TS structures can be located for the reaction, and the distinction between the BPC and DA mechanisms is blurred.
Using our enhanced sampling method, we aimed to describe the ``mechanistic continuum'' by implicitly taking into account the relevant TS structures.
By utilizing a pair of easily interpretable CVs, we obtained FESs in which the bifurcations were clearly visualized and consistent with previously located TS structures.

The effectiveness of a WTMetaD simulation depends substantially on the choice of CVs: they must act as low-dimensional descriptors that can discriminate the relevant metastable states and transition states.\cite{valsson2016,bussi2020}
Using the coordination numbers\cite{iannuzzi2003} given by
\begin{equation}
  c_i = \frac{1-(d_i/\sigma)^6}{1-(d_i/\sigma)^8} \qquad (i \in \{1,2,3\})
  \label{eqn:contact}
\end{equation}
where $d_i$ is the distance between two atoms as defined in Figure~\ref{fig:spnf}c and $\sigma$ is a reference distance (1.7 \AA), we employed a set of two CVs, $\mathbf{s} = (s_1, s_2)$, where $s_1 = c_1 + c_2 + c_3$ follows the overall reaction progress and $s_2 = c_2 - c_3$ discriminates between the two products.

Although \textbf{3-A} is considered to be the dominant [6+4] product\cite{patel2016} among the possible products with different stereochemistry (Figure~\ref{fig:spnf}b), an extensive TS search by Medvedev et al.\cite{medvedev2017} suggested the existence of several low-energy TS structures that are stereochemically relevant to other products (\textbf{3-B}--\textbf{D}).
Accordingly, in our metadynamics simulations, we considered all the possible [6+4] products.
However, since the transitions between metastable reactant states (\textbf{1-A}--\textbf{D}) leading to such products are hindered by the barriers between \textit{s-cis} and \textit{s-trans} isomers, using only the aforementioned CVs would require excessively long simulations.\cite{pietrucci2017}
To overcome this problem, we introduced a set of conformational restraints $W_\text{X}(\mathbf{R})$ ($\text{X} \in \{\mathrm{A,B,C,D}\}$) to constrain the substrate from accessing states other than \textbf{1-X}, \textbf{2}, and \textbf{3-X}, and independent WTMetaD simulations with each of the restraints.

From the WTMetaD simulations, the FES can be reconstructed up to a constant $C_\mathrm{X}$ by reweighting\cite{branduardi2012}:
\begin{align}
  F_\mathrm{X}(\mathbf{s})
   & = -\frac{1}{\beta} \log \left< \delta\left[\mathbf{s} - \mathbf{s}(\mathbf{R})\right] \right> + C_\mathrm{X} \nonumber                                                                                                              \\
   & = -\frac{1}{\beta} \log \frac{\left< \delta\left[\mathbf{s} - \mathbf{s}(\mathbf{R})\right] e^{+\beta V(\mathbf{s}(\mathbf{R}))} \right>_\mathrm{X}}{\left< e^{+\beta V(\mathbf{s}(\mathbf{R}))} \right>_\mathrm{X}} + C_\mathrm{X}
  \label{eqn:fes_wtmetad}
\end{align}
where $\left<\cdot\right>_\mathrm{X}$ denotes an average over the biased ensemble constructed from the WTMetaD run with restraint $W_\mathrm{X}(\mathbf{R})$.
To level off the FES from each WTMetaD run, we set the entire conformations of \textbf{1} as the reference state (i.e., $F_\mathbf{1} = 0$).
Then, we introduced another WTMetaD run (run T) with a set of three torsional angle CVs, $\Phi = (\phi_1, \phi_2, \phi_3)$ (as defined in Figure~\ref{fig:spnf}c), to effectively sample the conformations of \textbf{1} over the \textit{s-cis}/\textit{s-trans} barriers.
The free energy of state \textbf{1-X} ($\mathrm{X} \in \{\mathrm{A,B,C,D}\}$) can be calculated using the Zwanzig equation\cite{zwanzig1954} as
\begin{equation}
  e^{-\beta F_{\text{\textbf{1-X}}}} = \left< e^{-\beta W_\mathrm{X}(\mathbf{R})} \right>_\mathbf{1}
  = \frac{\left< e^{-\beta W_\mathrm{X}(\mathbf{R})} e^{+\beta V(\Phi(\mathbf{R}))} \right>_\mathrm{T}}{\left< e^{+\beta V(\Phi(\mathbf{R}))} \right>_\mathrm{T}}
  \label{eqn:zwanzig}
\end{equation}
where $\left<\cdot\right>_\mathbf{1}$ denotes an ensemble average over state \textbf{1} and $\left<\cdot\right>_\mathrm{T}$ a biased ensemble average obtained from WTMetaD run T.
Finally, $C_\mathrm{X}$ in eq~\ref{eqn:fes_wtmetad} can be determined by equating the free energy in eq~\ref{eqn:zwanzig} with the integral of the FES over $\Omega_{\text{\textbf{1-X}}}$, the CV space corresponding to \textbf{1-X}:

\begin{equation}
  e^{-\beta F_{\text{\textbf{1-X}}}} = \int_{\Omega_{\text{\textbf{1-X}}}} e^{-\beta F_\mathrm{X}(\mathbf{s})} \, \mathrm{d}\mathbf{s}
  \label{eqn:fes_shift}
\end{equation}

Initially, 1.8-ns WTMetaD simulations were run using a semiempirical tight-binding Hamiltonian, GFN-xTB.\cite{grimme2017}
The computational cost of the semiempirical Hamiltonian was low enough to run sufficiently long simulations to sample the configuration space; however, the method does not provide a quantitatively accurate description of the chemical process of interest.
Therefore, we employed the FEP scheme\cite{li2018,piccini2019} to refine the FES at the DFT level of theory [M06-2X/6-31G(d)\cite{zhao2008,ditchfield1971,hehre1972,hariharan1973}].
The ensemble average of an observable $O(\mathbf{R})$ at a high-level Hamiltonian $U_H(\mathbf{R})$ can be calculated from ensemble averages using a low-level Hamiltonian $U_L(\mathbf{R})$ by incorporating the perturbative weight:
\begin{equation}
  \left<O(\mathbf{R})\right>_H = \frac{\left<O(\mathbf{R}) e^{+\beta[U_L(\mathbf{R})-U_H(\mathbf{R})]}\right>_L}{\left< e^{+\beta[U_L(\mathbf{R})-U_H(\mathbf{R})]}\right>_L}
  \label{eqn:fep}
\end{equation}
where $\left<\cdot\right>_H$ and $\left<\cdot\right>_L$ are ensemble averages obtained using high-level and low-level Hamiltonians, respectively.
Accordingly, the biased ensemble averages on the right-hand sides of eqs~\ref{eqn:fes_wtmetad} and \ref{eqn:zwanzig} are refined using eq~\ref{eqn:fep}.

\begin{figure}
  \includegraphics[width=\linewidth]{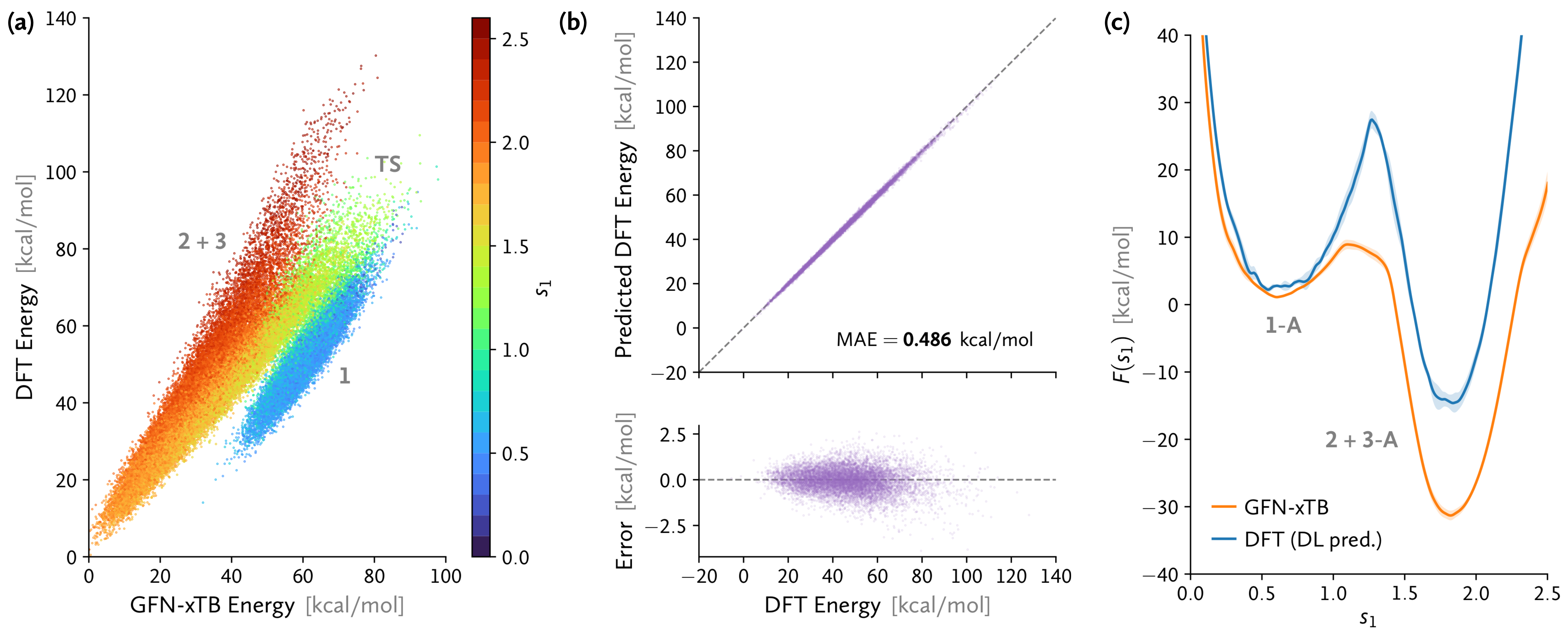}
  \caption{(a) GFN-xTB and DFT energies at the M06-2X/6-31G(d) level of theory for the configurations sampled in the well-tempered metadynamics simulations used as the training set.
  The points are colored according to the reaction progress, as represented by the CV $s_1$.
  (b) Comparison of the actual DFT energies and the energies predicted by the graph neural network for the test set configurations (MAE: mean absolute error).
  (c) One-dimensional free energy profiles along the reaction progress obtained at GFN-xTB and DFT levels as calculated from the energies predicted by the deep learning model.
  Shaded areas represent the standard deviation from five independent simulations.}
  \label{fig:xtb2dft}
\end{figure}

Previous methods based on FEP\cite{piccini2019,raucci2022} sampled a small number of configurations to calculate the DFT energies and perturbative weights and then applied post-processing such as moving average calculations to smooth the obtained high-level FES.
As an alternative approach, in this study we utilized a graph neural network method inspired by molecular mechanics\cite{zhang2020mxm} trained on the DFT calculation results for a small subset of the configurations sampled from WTMetaD runs to estimate the DFT energies of a large number of configurations.
The network was trained to simultaneously predict the differences between DFT and GFN-xTB energies, solvation energies in implicit water (CPCM\cite{barone1998,cossi2003}), and CM5 partial atomic charges\cite{marenich2012}.
As well as offering additional information for the system, such as the effect of implicit solvent, the multi-task setting\cite{caruana1997} also provides more accurate DFT energy predictions (see Table~S2).
As shown in Figure~\ref{fig:xtb2dft}a, the GFN-xTB and DFT energies are correlated but quantitatively different, and the values are also dependent on the reaction progress (CV $s_1$).
The constructed deep learning model can accurately predict the DFT energy with a mean absolute error (MAE) of 0.486 kcal/mol on the test set (Figure~\ref{fig:xtb2dft}b).
Furthermore, the prediction error is less than 2 kcal/mol for more than 99\% of the test set configurations, and the magnitude of the error does not depend on the reaction progress (see Figure~S4).
Hence, the deep learning model can be used as a reliable surrogate for the high-cost DFT single-point energy calculations, and when applied to FEP calculations, the resulting FES reflects the relatively high TS and product energies obtained from the DFT-level calculations (Figure~\ref{fig:xtb2dft}c).
Details of the restraints, WTMetaD simulations, and deep learning models are reported in the \hyperref[sec:methods]{Computational Methods} section.


\begin{figure}
  \includegraphics[width=\linewidth]{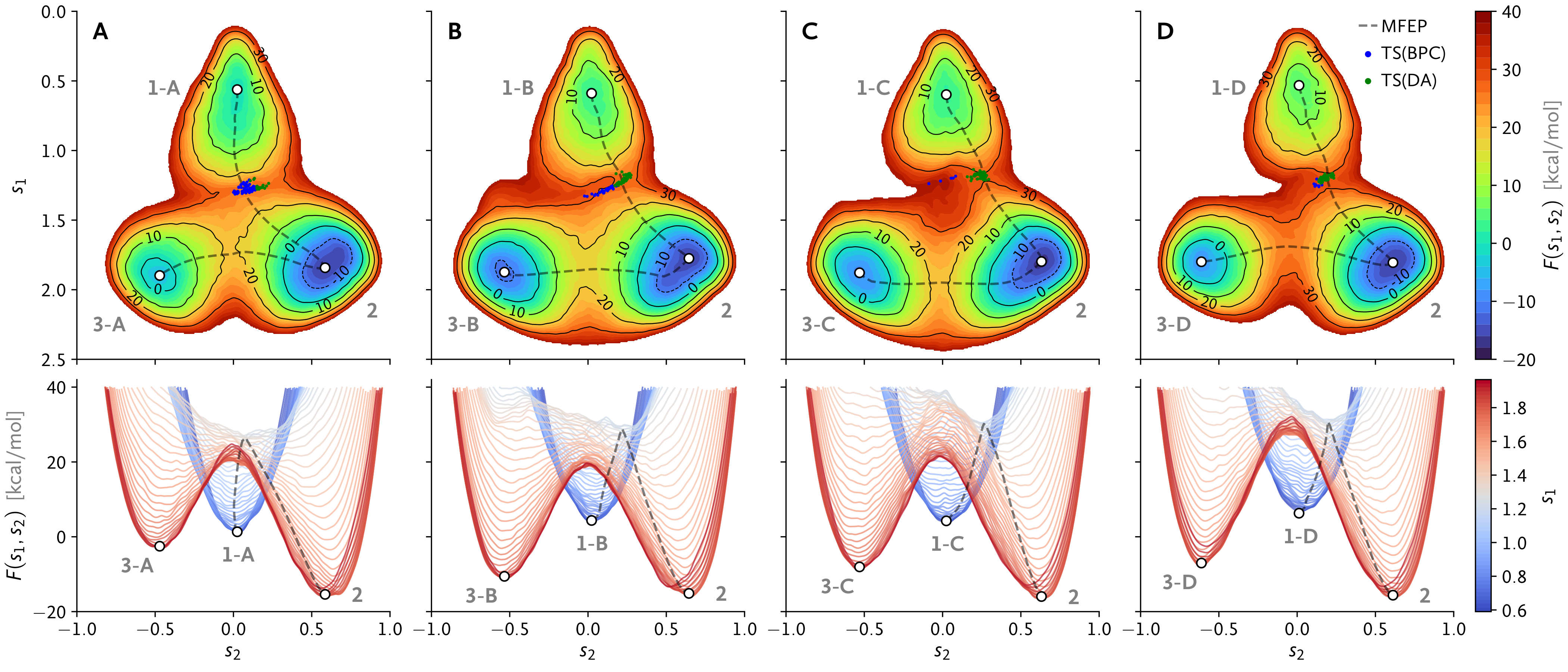}
  \caption{Upper panels: two-dimensional free energy surfaces, $F(s_1, s_2)$, obtained from the well-tempered metadynamics simulations with different stereochemical restraints (A--D).
  The simulations were run using the GFN-xTB Hamiltonian and the resulting free energy surfaces were refined at the M06-2X/6-31G(d) level of theory using the single-point energies predicted by the graph neural network.
  Minimum free energy paths (MFEPs) between the free energy minima and CV-space projections of the transition state structures from ref~\citenum{medvedev2017}, re-optimized at the current level of theory, are also shown.
  Lower panels: cross-sections of the free energy surfaces along the reaction progress coordinate (CV $s_1$).}
    \label{fig:fes2d}
\end{figure}

Two-dimensional free energy profiles, $F(s_1, s_2)$, obtained from each WTMetaD run (A, B, C, and D) with FEP refinement using eqs~\ref{eqn:fes_wtmetad}, \ref{eqn:zwanzig}, and \ref{eqn:fep} are shown in Figure~\ref{fig:fes2d}.
The free energy surfaces provide a quantitative description of the asymmetric post-TS bifurcations, and cross-sections along the reaction progress (CV $s_1$) illustrate the valley--ridge inflection point on the FES.
The free energy landscapes near the transition region, especially the skewness relative to $s_2 = 0$ and the activation barriers, depend on the stereochemical pathways (A--D) determined by the respective stereochemical restraints.
To better understand the product selectivity, we calculated the minimum free energy paths (MFEPs) using the nudged elastic band (NEB) algorithm.\cite{henkelman2000}
MFEPs between the free energy minima of the reactants and products show that the pre-TS pathway for A aligns well with the $s_2 = 0$ line and the transition region has more ``BPC-like'' character, whereas the pathways for B--D are tilted toward the [4+2] product (\textbf{2}, $s_2 > 0$) and the transition region is ``DA-like.''
Although the MFEPs for B--D are similar, the curvatures of the FESs near the TS region are different, and a two-dimensional description is required to account for the product selectivity.
We also obtained the TS structures at the current level of theory, by re-optimizing the TS structures reported by Medvedev et al.,\cite{medvedev2017} and projected them onto the CV space of the relevant stereochemistry (A--D).
From Figure~\ref{fig:fes2d} and the TS energies given in Figure~S3, it can be seen that free energy landscapes and associated MFEPs from the WTMetaD simulations (and FEP refinement) represent the energetic landscapes of the searched TS structures well.
This result indicates that the free energy landscapes, reaction energetics, and preferred reaction pathways and mechanisms of the system considered were efficiently characterized using our methodology, obviating the need for an extensive manual search for TS structures.

\begin{figure}
  \includegraphics[width=0.5\linewidth]{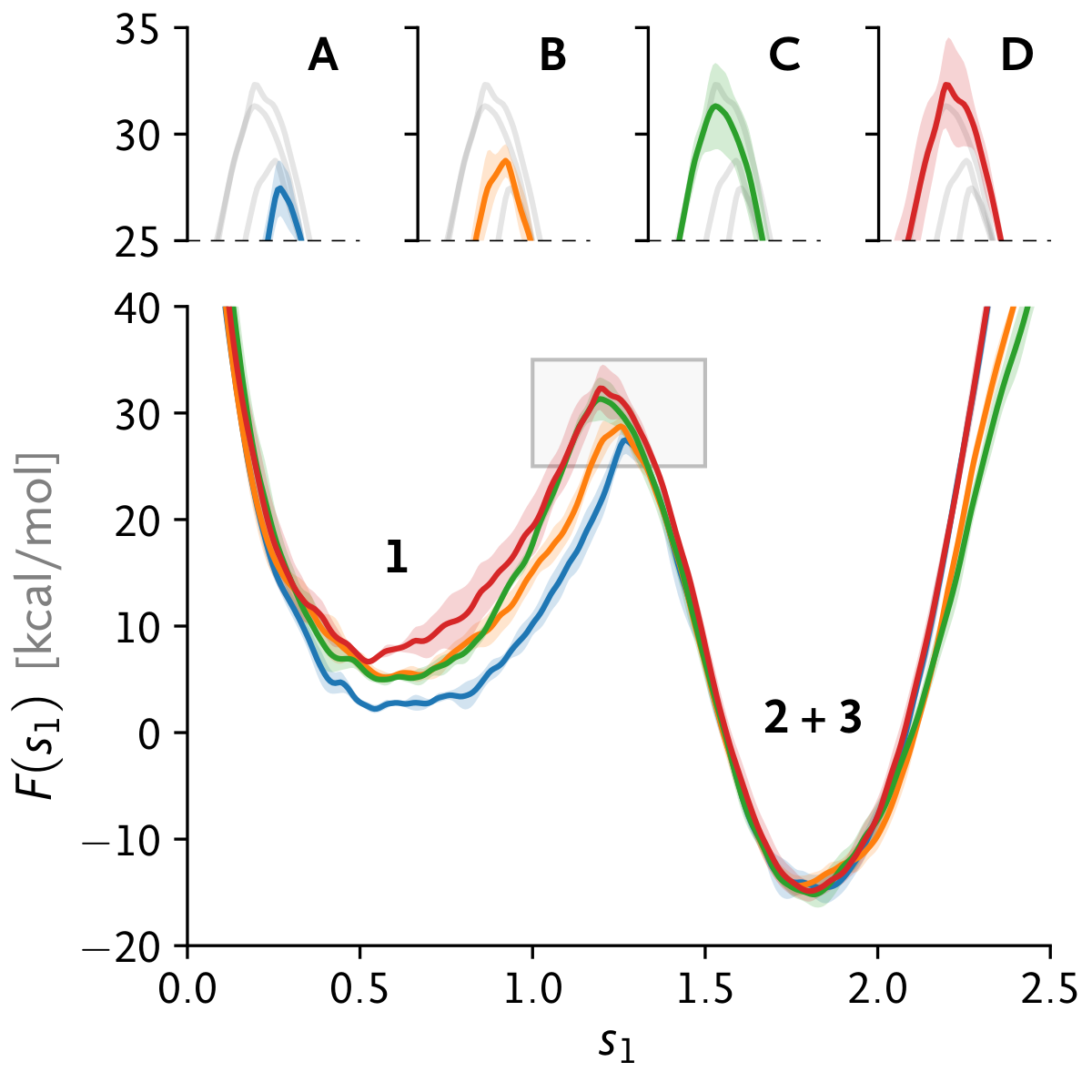}
  \caption{One-dimensional profiles of the free energy $F$ versus the reaction progress coordinate $s_1$ obtained from the well-tempered metadynamics simulations with different stereochemical restraints (A--D).
  Expanded views of the transition region (indicated by the overlaid box in the lower panel) are shown in the upper panels to highlight the differences between the activation barriers.
  Shaded areas represent the standard deviations from five independent simulations.}
    \label{fig:fes1d}
\end{figure}

In order to investigate the differences between the activation barriers of the different stereochemical pathways, one-dimensional profiles of the free energy versus the reaction progress were calculated (Figure~\ref{fig:fes1d}).
The results indicate that pathway A, with the lowest barrier, is the dominant stereochemical pathway, and the barrier increases in the order $\text{A} < \text{B} \ll \text{C} < \text{D}$.
As the barrier height depends on the choice of CVs, the gauge correction scheme described by Bal et al.\cite{bal2020} should be introduced for consistency.
The free energy barrier $\Delta^\ddagger F$ is given as
\begin{equation}
  \Delta^\ddagger F = F_\text{TS}^G + \frac{1}{\beta} \log \frac{\lambda}{h} \sqrt{\frac{2\pi m}{\beta}}
  \label{eqn:barrier}
\end{equation}
with the gauge-invariant geometric FES\cite{hartmann2007} $F^G(s_1)$ defined as
\begin{equation}
  F^G(s_1) = F(s_1) - \frac{1}{\beta} \log \left< \lambda \left\vert \nabla_\mathbf{R} s_1 \right\vert \right>_{s_1}
  \label{eqn:geomfes}
\end{equation}
where $\lambda$ is the length unit for the system coordinates $\mathbf{R}$ and $m$ is the reduced mass for the pair of atoms defining the CVs.
The free energies of states \textbf{1-X} (eq~\ref{eqn:zwanzig}) and the free energy barriers $\Delta^\ddagger F_\text{X}$ (eq~\ref{eqn:barrier}) for the respective stereochemical pathways ($\text{X} \in \{\mathrm{A,B,C,D}\}$) are reported in Table~\ref{table:barrier}.

\begin{table}
  \caption{Free Energies of the Reactant States and Activation Free Energies for Different Stereochemical Pathways\textsuperscript{\emph{a}}}
  \label{table:barrier}
  \begin{tabular}{ccc}
    \hline
    Pathway (X) & $F_{\text{\textbf{1-X}}}$ (kcal/mol) & $\Delta^\ddagger F_\text{X}$ (kcal/mol) \\
    \hline
    A & $3.2 \pm 0.6$ & $28.3 \pm 1.4$ \\
    B & $6.1 \pm 1.1$ & $29.6 \pm 1.4$ \\
    C & $5.8 \pm 0.8$ & $32.1 \pm 2.1$ \\
    D & $8.0 \pm 1.8$ & $33.2 \pm 2.7$ \\
    \hline \\
  \end{tabular}

  \textsuperscript{\emph{a}}The free energies of the reactant states are reported with the standard deviation from five independent simulations. The uncertainty was propagated to the activation free energies since they were calculated from the FES shifted using $F_{\text{\textbf{1-X}}}$ (eq~\ref{eqn:fes_shift}).
\end{table}

This result quantifies the aforementioned trend in the activation barriers of the different stereochemical pathways, with the lowest activation free energy, $\Delta^\ddagger F_\text{A}$, being 28.3 kcal/mol.
Only small changes in the reaction energetics were obtained when the solvation energy in implicit water was also applied at the FEP refining step (Table~S1), with $\Delta^\ddagger F_\text{A}$ becoming 28.2 kcal/mol.
Despite this activation barrier being similar to the value obtained using harmonic transition state theory and the same DFT functional (27.6 kcal/mol),\cite{patel2016} it is higher than the experimental activation barrier of 22.0 kcal/mol in water.\cite{kim2011}
As the relative stability of configurations can be affected by hydrogen bonding,\cite{yang2018a,chen2018} QM/MM simulations are required for a more accurate description of the reaction.

\begin{figure}
  \includegraphics[width=0.6\linewidth]{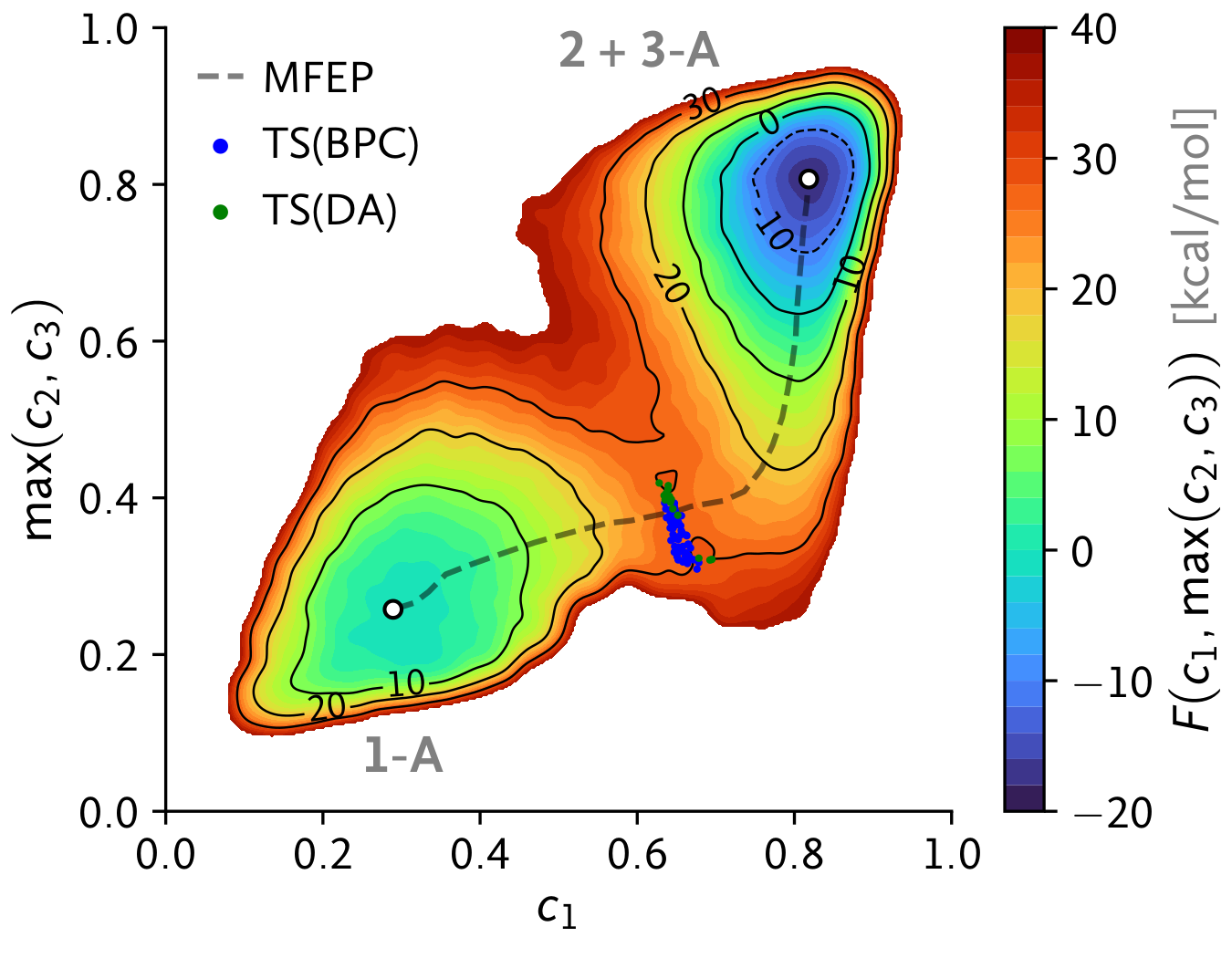}
  \caption{Two-dimensional free energy profile (More O'Ferrall--Jencks plot) calculated from the well-tempered metadynamics simulation A.
  The results for the other stereochemical pathways (B--D) are similar and are shown in Figure~S2.
  The formation of the two bonds is represented by the coordination number CVs, $c_1$ and $\max(c_2, c_3)$ (see Figure~\ref{fig:spnf}c and eq~\ref{eqn:contact}).
  The minimum free energy path (MFEP) between the free energy minima and the transition state structures are displayed in the same way as those in Figure~\ref{fig:fes2d}.}
    \label{fig:mofj}
\end{figure}

Previous quasi-classical dynamics studies\cite{patel2016,yang2018a} have found that the [4+2]/[6+4] cycloadditions that are the focus of our study are energetically concerted but are characterized by stepwise dynamics.
This has also been verified by means of kinetic isotope effect experiments.\cite{jeon2017}
To determine the step-wise bond formation dynamics, we constructed the FES with another set of CVs: $c_1$ and $\max(c_2, c_3)$.
Since these two CVs follow the formation of the first and second bonds during the cycloadditions, the resulting FES can be regarded as a More O'Ferrall--Jencks plot.\cite{anslyn2006}
The obtained FES, the MFEP between the free energy minima, and the projected TS locations for stereochemical pathway A are shown in Figure~\ref{fig:mofj}.
The reaction pathway, as represented by the MFEP, is curved toward the bottom right corner and indicates the two bonds are formed asynchronously.
The free energy maximum along the reaction pathway coincides with the projected TS positions and corresponds to a substrate geometry in which the first bond is almost formed and the second is not.
Similar results were obtained for the other stereochemical pathways (B--D), and these are reported in Figure~S2.


We proposed a method to explore DFT-level free energy landscapes with post-TS bifurcation, using WTMetaD simulation and FEP refinement assisted by a deep learning energy estimator.
Using our methodology, it is possible to characterize a bifurcating reaction pathway in a manner independent of a specific TS structure, as was demonstrated for the case of a SpnF-catalyzed Diels--Alder reaction.
In addition, we showed that deep learning methods, especially graph neural networks, can be a reliable surrogate for costly single-point DFT calculations for FEP refining steps.
In future studies, it would be advantageous to make use of specialized CVs\cite{bonati2020,trizio2021,bonati2021} and recently developed enhanced sampling methods\cite{invernizzi2020,invernizzi2022} to study more complex cases (e.g., systems with post-TS trifurcations\cite{jamieson2021}).
Furthermore, to obtain a more accurate description of the reaction rates and product selectivity in these systems, future studies may benefit from the incorporation of explicit kinetics modeling.\cite{tiwary2013,mccarty2015}

\section{Computational Methods} \label{sec:methods}

\textbf{Conformational restraints.}
While no restraint was applied to WTMetaD run T, two types of restraint were used for the other WTMetaD runs (A, B, C, D): one to prevent unwanted reactions, and the other to restrict the reactant to the stereochemical pathway of interest.

We defined the coordination number $c_{ij} \in [0, 1]$, angle $\theta_{ijkl} \in [0, \pi]$, and torsional angle $\phi_{ijkl} \in (-\pi, \pi]$ as follows:
\begin{align}
  c_{ij} &= \frac{1-(\vert\mathbf{r}_{ij}\vert/\sigma_{ij})^6}{1-(\vert\mathbf{r}_{ij}\vert/\sigma_{ij})^8}
  \label{eqn:contact_alt} \\
  \theta_{ijkl} &= \cos^{-1} \left( \frac{\mathbf{r}_{ji} \cdot \mathbf{r}_{kl}}{\left\vert\mathbf{r}_{ji}\right\vert \left\vert\mathbf{r}_{kl}\right\vert} \right)
  \label{eqn:angle} \\
  \phi_{ijkl} &= \operatorname{atan2} \left( \left\vert \mathbf{r}_{jk} \right\vert \mathbf{r}_{ij} \cdot (\mathbf{r}_{jk} \times \mathbf{r}_{kl}), (\mathbf{r}_{ij} \times \mathbf{r}_{jk}) \cdot (\mathbf{r}_{jk} \times \mathbf{r}_{kl}) \right)
  \label{eqn:torsion}
\end{align}
where $\sigma_{ij}$ is 1.7 \AA{} and 1.6 \AA{} for the C--C and C--O atom pairs, respectively.
First, to avoid the sampling of irrelevant reaction pathways, we restrained four atom pairs, $\mathcal{P} = \left\{ (2, 11), (3, 14), (1, 13), (11, 19) \right\}$, from being too close using the harmonic wall restraint
\begin{equation}
  W_\text{reac} = k \sum_{(i,j)\in\mathcal{P}} \max(c_{ij}-0.4,\, 0)^2
  \label{eqn:res_reac}
\end{equation}
with $k = 5 \times 10^3$ kcal/mol.
Further, to constrain the molecule to exploration of the relevant stereochemical pathway ($\text{X} \in \{\mathrm{A,B,C,D}\}$), we also introduced the following harmonic wall restraint:
\begin{equation}
  W_\text{stereo,X} = \kappa \sum_{i=1}^{10} \max(\alpha_{i,\text{X}} \cos \varphi_i,\, 0)^2
  \label{eqn:res_stereo}
\end{equation}
with $\kappa = 50$ kcal/mol and the coefficients $\alpha_{i,\text{X}}$ and angles $\varphi_i$ defined as in Table~\ref{table:indices}; $\alpha_{i,\text{X}}$ values of $-1$ and $+1$ restrain the angle $\phi_i$ to values corresponding to the \textit{syn} and \textit{anti} configurations, respectively;$\varphi_1$ to $\varphi_7$ determine stereochemistry of the [4+2] product (\textbf{2}); and $\varphi_8$ to $\varphi_{10}$ control the stereochemical pathway leading to the [6+4] product (\textbf{3}).

\begin{table}
  \caption{Coefficients and Angles Used to Define the Restraints and Corresponding Substrate Atom Indices}
  \label{table:indices}
  \includegraphics[width=0.5\linewidth]{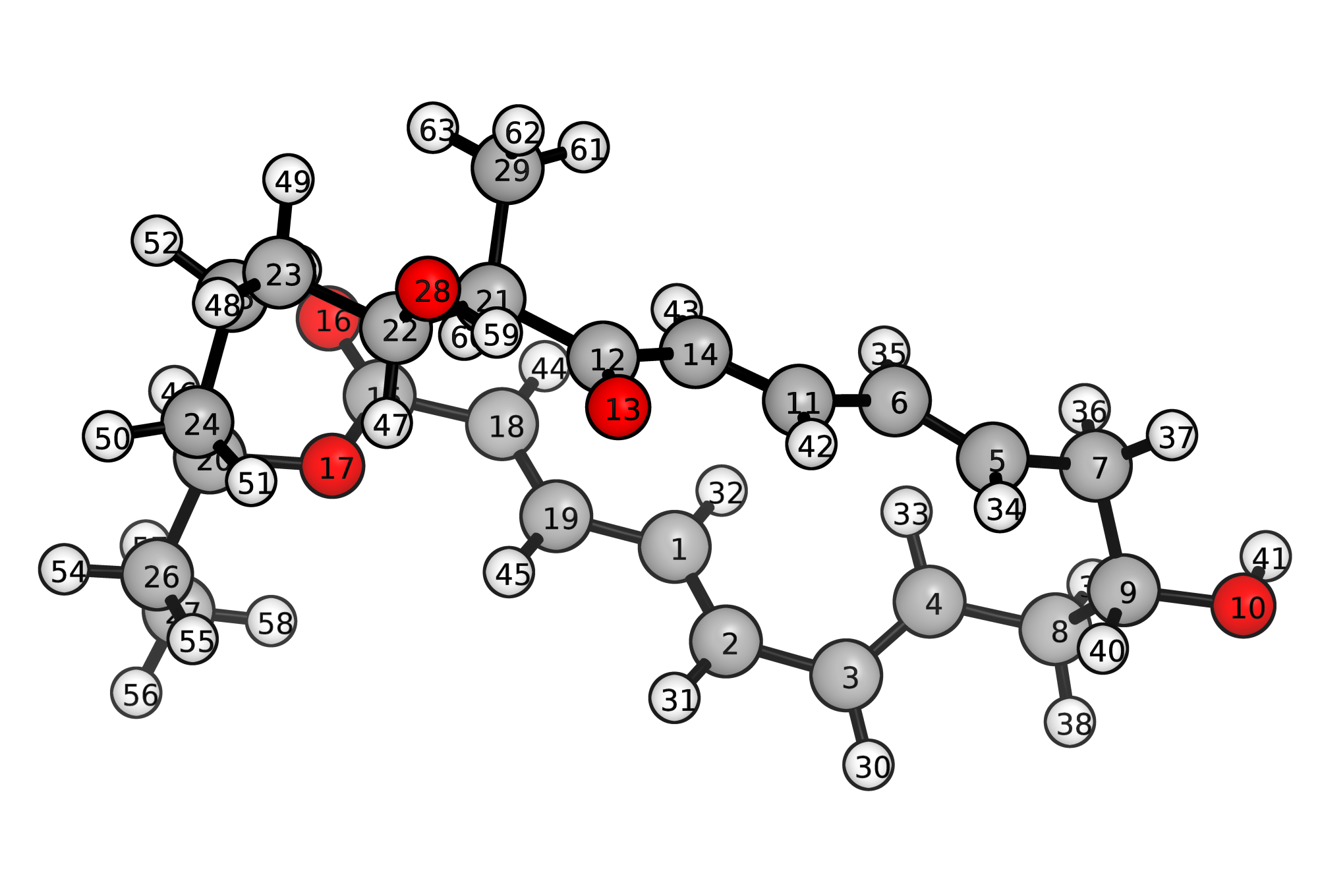}
  \begin{tabular}{c|c|cccc}
  \hline
  $i$ & $\varphi_i$ & $\alpha_{i,\text{A}}$ & $\alpha_{i,\text{B}}$ & $\alpha_{i,\text{C}}$ & $\alpha_{i,\text{D}}$ \\ \hline
  1   & $\phi_{34,5,9,40}$     & $-1$  & $-1$  & $-1$  & $-1$  \\
  2   & $\phi_{33,4,9,40}$     & $+1$  & $+1$  & $+1$  & $+1$  \\
  3   & $\phi_{35,6,5,34}$     & $+1$  & $+1$  & $+1$  & $+1$  \\
  4   & $\phi_{32,1,4,33}$     & $-1$  & $-1$  & $-1$  & $-1$  \\
  5   & $\phi_{1,2,3,4}$       & $-1$  & $-1$  & $-1$  & $-1$  \\
  6   & $\theta_{34,5,4,33}$   & $+1$  & $+1$  & $+1$  & $+1$  \\
  7   & $\theta_{35,6,1,32}$   & $-1$  & $-1$  & $-1$  & $-1$  \\
  8   & $\phi_{43,14,6,35}$    & $-1$  & $-1$  & $+1$  & $+1$  \\
  9   & $\phi_{44,18,1,32}$    & $-1$  & $+1$  & $-1$  & $+1$  \\
  10  & $\theta_{43,14,18,44}$ & $-1$  & $+1$  & $+1$  & $-1$  \\
  \hline
  \end{tabular}
\end{table}

Then, the restraints for the WTMetaD runs, mentioned in the main text, are defined by $W_\text{X}(\mathbf{R}) = W_\text{reac} + W_\text{stereo,X}$.
Although the stereochemical restraint may prevent the sampling of the full conformational space of the products, its effect on the obtained FES in this work is insignificant, and this is apparent from the consistency of the FESs in the product region in Figure~\ref{fig:fes1d}.

\textbf{Simulation settings.}
WTMetaD simulations were carried out using the CP2K 9.1 software package\cite{cp2k} patched with PLUMED 2.7.3.\cite{plumed2}
The semiempirical GFN-xTB Hamiltonian\cite{grimme2017} was used to obtain trajectories with an integration time step of 0.5 fs.
A canonical ($NVT$) ensemble at 298.15 K was sampled using the velocity rescaling thermostat\cite{bussi2007} with a time constant of 50 fs.
For each WTMetaD run (A, B, C, D, or T) type, the initial structure was obtained by energy minimization at the GFN-xTB level.
The system was equilibrated under the application of the respective constraint $W_\text{X}(\mathbf{R})$ for 1.0 ns, and five configurations (positions and velocities) were sampled from the last 0.8 ns to initiate five statistically independent WTMetaD runs for each type.
For WTMetaD runs A, B, C, and D, a bias factor ($\gamma$) of 30 was used, and Gaussian biases with a height of 1.0 kcal/mol and width of 0.05 were added every 200 MD steps.
For WTMetaD run T, a bias factor of 15 was used, and Gaussian biases with a height of 0.5 kcal/mol and a width of 0.2 rad were added every 200 MD steps.
All the WTMetaD simulations were run for 1.8 ns, and the last 1.2 ns was used to obtain the free energy profiles.

\textbf{DFT calculations.}
The single point energy calculations for the sampled configurations and the TS structure optimizations were carried out at the M06-2X/6-31G(d) level of theory\cite{zhao2008,ditchfield1971,hehre1972,hariharan1973} as implemented in Gaussian 16, revision C.01.\cite{g16}
The keywords used to designate the level of theory and implicit solvation were \texttt{m062x/6-31g(d)} and \texttt{scrf=(cpcm,solvent=water)}, respectively.
We used the keyword \texttt{pop=hirshfeld} to obtain CM5 partial charges.
The transition states were optimized using the keyword \texttt{opt=(ts,calcfc,noeigen)}.

The transition state structures in ref~\citenum{medvedev2017} were obtained at the M06-2X/6-31+G(d) level of theory.
We re-optimized these transition state structures at our level of theory [M06-2X/6-31G(d)] to verify that the transition regions in the free energy landscapes obtained in our work were consistent with the results of harmonic transition state theory.
With the lowest energy TS as a reference, only TSs with energies of less than 15 kcal/mol were used.
The projected TS locations are shown in the upper panels of Figure~\ref{fig:fes2d}, and the relative energies are reported in Figure~S3.

\textbf{Deep learning model.}
The geometric deep learning model used in this work was MXMNet.\cite{zhang2020mxm}
While other similar models (e.g., DimeNet\textsuperscript{++}\cite{klicpera2020}) might have been considered suitable for our task, we did not benchmark the performance of other models as DFT-level energy estimation by deep learning was not the main source of error for the free energy surfaces obtained in this work.

The graph neural network used in this study simultaneously predicted (1) the differences between DFT and GFN-xTB energies ($E_\text{DFT} - E_\text{xTB}$), (2) the solvation energies in implicit water ($E_\text{DFT/solv} - E_\text{DFT}$), and (3) the CM5 partial atomic charges ($\mathbf{q}$).
Accordingly, the MXMNet architecture was modified so that each Local Layer Message Passing Module had three separate ``Output'' layers, one for each property to be predicted (Figure~3c in ref~\citenum{zhang2020mxm}).
The target values for the energies were obtained by subtracting the reference values obtained from the molecular geometry with the lowest $E_\text{DFT}$.
Energy value units of kcal/mol were used, along with partial charge units of $e$.

The deep learning model was optimized by minimizing the weighted sum of the L1 losses for the targets, i.e.,
\begin{equation}
  \mathcal{L} = L_1(E_\text{DFT} - E_\text{xTB}) + w_s \cdot L_1(E_\text{DFT/solv} - E_\text{DFT}) + w_c \cdot L_1(\mathbf{q})
  \label{eqn:loss}
\end{equation}
where $L_1(\cdot)$ denotes the mean absolute error between the predicted and true values.
Several values for the weights $w_s$ and $w_c$ were tested as in Table~S2, and the weights for the final ensemble were determined as $w_s = 1$ and $w_c = 2500$.
The loss was optimized by the Adam optimizer\cite{kingma2014} with an initial learning rate of $3 \times 10^{-4}$.
The learning rate was warmed up linearly from zero to the initial learning rate for the first five epochs, then decayed exponentially by a factor of $0.5^{1/250}$ every epoch.
We used a batch size of 64 and a hidden dimension size of 128; the number of layers was 6, the gradient clipping norm was 100, and the global and local layer distance cutoff values were 10 \AA\ and 5 \AA, respectively.

\begin{acknowledgement}
The authors thank Jaehak Lee for their careful reading of the manuscript and constructive suggestions.
\end{acknowledgement}

\begin{suppinfo}
Free energy profiles calculated at the GFN-xTB level using different stereochemical restraints, with corresponding transition state energies; calculated free energies with implicit solvation; error distribution of the deep learning model outputs
\end{suppinfo}

\bibliography{main}

\providecommand{\latin}[1]{#1}
\makeatletter
\providecommand{\doi}
  {\begingroup\let\do\@makeother\dospecials
  \catcode`\{=1 \catcode`\}=2 \doi@aux}
\providecommand{\doi@aux}[1]{\endgroup\texttt{#1}}
\makeatother
\providecommand*\mcitethebibliography{\thebibliography}
\csname @ifundefined\endcsname{endmcitethebibliography}
  {\let\endmcitethebibliography\endthebibliography}{}
\begin{mcitethebibliography}{81}
\providecommand*\natexlab[1]{#1}
\providecommand*\mciteSetBstSublistMode[1]{}
\providecommand*\mciteSetBstMaxWidthForm[2]{}
\providecommand*\mciteBstWouldAddEndPuncttrue
  {\def\EndOfBibitem{\unskip.}}
\providecommand*\mciteBstWouldAddEndPunctfalse
  {\let\EndOfBibitem\relax}
\providecommand*\mciteSetBstMidEndSepPunct[3]{}
\providecommand*\mciteSetBstSublistLabelBeginEnd[3]{}
\providecommand*\EndOfBibitem{}
\mciteSetBstSublistMode{f}
\mciteSetBstMaxWidthForm{subitem}{(\alph{mcitesubitemcount})}
\mciteSetBstSublistLabelBeginEnd
  {\mcitemaxwidthsubitemform\space}
  {\relax}
  {\relax}

\bibitem[Ess \latin{et~al.}(2008)Ess, Wheeler, Iafe, Xu,
  {\c{C}}elebi-{\"O}l{\c{c}}{\"u}m, and Houk]{ess2008}
Ess,~D.~H.; Wheeler,~S.~E.; Iafe,~R.~G.; Xu,~L.;
  {\c{C}}elebi-{\"O}l{\c{c}}{\"u}m,~N.; Houk,~K.~N. {Bifurcations on Potential
  Energy Surfaces of Organic Reactions}. \emph{Angew. Chem., Int. Ed.}
  \textbf{2008}, \emph{47}, 7592--7601\relax
\mciteBstWouldAddEndPuncttrue
\mciteSetBstMidEndSepPunct{\mcitedefaultmidpunct}
{\mcitedefaultendpunct}{\mcitedefaultseppunct}\relax
\EndOfBibitem
\bibitem[Rehbein and Carpenter(2011)Rehbein, and Carpenter]{rehbein2011}
Rehbein,~J.; Carpenter,~B.~K. {Do We Fully Understand What Controls Chemical
  Selectivity?} \emph{Phys. Chem. Chem. Phys.} \textbf{2011}, \emph{13},
  20906--20922\relax
\mciteBstWouldAddEndPuncttrue
\mciteSetBstMidEndSepPunct{\mcitedefaultmidpunct}
{\mcitedefaultendpunct}{\mcitedefaultseppunct}\relax
\EndOfBibitem
\bibitem[Hare and Tantillo(2017)Hare, and Tantillo]{hare2017a}
Hare,~S.~R.; Tantillo,~D.~J. {Post-Transition State Bifurcations Gain Momentum
  -- Current State of the Field}. \emph{Pure Appl. Chem.} \textbf{2017},
  \emph{89}, 679--698\relax
\mciteBstWouldAddEndPuncttrue
\mciteSetBstMidEndSepPunct{\mcitedefaultmidpunct}
{\mcitedefaultendpunct}{\mcitedefaultseppunct}\relax
\EndOfBibitem
\bibitem[Yamataka \latin{et~al.}(1999)Yamataka, Aida, and Dupuis]{yamataka1999}
Yamataka,~H.; Aida,~M.; Dupuis,~M. One Transition State Leading to Two Product
  States: Ab Initio Molecular Dynamics Simulations of the Reaction of
  Formaldehyde Radical Anion and Methyl Chloride. \emph{Chem. Phys. Lett.}
  \textbf{1999}, \emph{300}, 583--587\relax
\mciteBstWouldAddEndPuncttrue
\mciteSetBstMidEndSepPunct{\mcitedefaultmidpunct}
{\mcitedefaultendpunct}{\mcitedefaultseppunct}\relax
\EndOfBibitem
\bibitem[Caramella \latin{et~al.}(2002)Caramella, Quadrelli, and
  Toma]{caramella2002}
Caramella,~P.; Quadrelli,~P.; Toma,~L. An Unexpected Bispericyclic Transition
  Structure Leading to 4+2 and 2+4 Cycloadducts in the \textit{Endo}
  Dimerization of Cyclopentadiene. \emph{J. Am. Chem. Soc.} \textbf{2002},
  \emph{124}, 1130--1131\relax
\mciteBstWouldAddEndPuncttrue
\mciteSetBstMidEndSepPunct{\mcitedefaultmidpunct}
{\mcitedefaultendpunct}{\mcitedefaultseppunct}\relax
\EndOfBibitem
\bibitem[Ussing \latin{et~al.}(2006)Ussing, Hang, and Singleton]{ussing2006}
Ussing,~B.~R.; Hang,~C.; Singleton,~D.~A. Dynamic Effects on the
  Periselectivity, Rate, Isotope Effects, and Mechanism of Cycloadditions of
  Ketenes with Cyclopentadiene. \emph{J. Am. Chem. Soc.} \textbf{2006},
  \emph{128}, 7594--7607\relax
\mciteBstWouldAddEndPuncttrue
\mciteSetBstMidEndSepPunct{\mcitedefaultmidpunct}
{\mcitedefaultendpunct}{\mcitedefaultseppunct}\relax
\EndOfBibitem
\bibitem[{\c{C}}elebi-{\"O}l{\c{c}}{\"u}m
  \latin{et~al.}(2007){\c{C}}elebi-{\"O}l{\c{c}}{\"u}m, Ess, Aviyente, and
  Houk]{ccelebi2007}
{\c{C}}elebi-{\"O}l{\c{c}}{\"u}m,~N.; Ess,~D.~H.; Aviyente,~V.; Houk,~K. Lewis
  Acid Catalysis Alters the Shapes and Products of Bis-Pericyclic Diels--Alder
  Transition States. \emph{J. Am. Chem. Soc.} \textbf{2007}, \emph{129},
  4528--4529\relax
\mciteBstWouldAddEndPuncttrue
\mciteSetBstMidEndSepPunct{\mcitedefaultmidpunct}
{\mcitedefaultendpunct}{\mcitedefaultseppunct}\relax
\EndOfBibitem
\bibitem[Thomas \latin{et~al.}(2008)Thomas, Waas, Harmata, and
  Singleton]{thomas2008}
Thomas,~J.~B.; Waas,~J.~R.; Harmata,~M.; Singleton,~D.~A. Control Elements in
  Dynamically Determined Selectivity on a Bifurcating Surface. \emph{J. Am.
  Chem. Soc.} \textbf{2008}, \emph{130}, 14544--14555\relax
\mciteBstWouldAddEndPuncttrue
\mciteSetBstMidEndSepPunct{\mcitedefaultmidpunct}
{\mcitedefaultendpunct}{\mcitedefaultseppunct}\relax
\EndOfBibitem
\bibitem[Hong and Tantillo(2009)Hong, and Tantillo]{hong2009}
Hong,~Y.~J.; Tantillo,~D.~J. A Potential Energy Surface Bifurcation in Terpene
  Biosynthesis. \emph{Nat. Chem.} \textbf{2009}, \emph{1}, 384--389\relax
\mciteBstWouldAddEndPuncttrue
\mciteSetBstMidEndSepPunct{\mcitedefaultmidpunct}
{\mcitedefaultendpunct}{\mcitedefaultseppunct}\relax
\EndOfBibitem
\bibitem[Siebert \latin{et~al.}(2011)Siebert, Zhang, Addepalli, Tantillo, and
  Hase]{siebert2011}
Siebert,~M.~R.; Zhang,~J.; Addepalli,~S.~V.; Tantillo,~D.~J.; Hase,~W.~L. The
  Need for Enzymatic Steering in Abietic Acid Biosynthesis: Gas-Phase Chemical
  Dynamics Simulations of Carbocation Rearrangements on a Bifurcating Potential
  Energy Surface. \emph{J. Am. Chem. Soc.} \textbf{2011}, \emph{133},
  8335--8343\relax
\mciteBstWouldAddEndPuncttrue
\mciteSetBstMidEndSepPunct{\mcitedefaultmidpunct}
{\mcitedefaultendpunct}{\mcitedefaultseppunct}\relax
\EndOfBibitem
\bibitem[Zhang \latin{et~al.}(2015)Zhang, Wang, Yao, Wang, and Yu]{zhang2015}
Zhang,~L.; Wang,~Y.; Yao,~Z.-J.; Wang,~S.; Yu,~Z.-X. Kinetic or Dynamic Control
  on a Bifurcating Potential Energy Surface? An Experimental and DFT Study of
  Gold-Catalyzed Ring Expansion and Spirocyclization of
  2-Propargyl-$\beta$-Tetrahydrocarbolines. \emph{J. Am. Chem. Soc.}
  \textbf{2015}, \emph{137}, 13290--13300\relax
\mciteBstWouldAddEndPuncttrue
\mciteSetBstMidEndSepPunct{\mcitedefaultmidpunct}
{\mcitedefaultendpunct}{\mcitedefaultseppunct}\relax
\EndOfBibitem
\bibitem[Patel \latin{et~al.}(2016)Patel, Chen, Yang, Guti{\'e}rrez, Liu, Houk,
  and Singleton]{patel2016}
Patel,~A.; Chen,~Z.; Yang,~Z.; Guti{\'e}rrez,~O.; Liu,~H.-w.; Houk,~K.;
  Singleton,~D.~A. Dynamically Complex [6+4] and [4+2] Cycloadditions in the
  Biosynthesis of Spinosyn A. \emph{J. Am. Chem. Soc.} \textbf{2016},
  \emph{138}, 3631--3634\relax
\mciteBstWouldAddEndPuncttrue
\mciteSetBstMidEndSepPunct{\mcitedefaultmidpunct}
{\mcitedefaultendpunct}{\mcitedefaultseppunct}\relax
\EndOfBibitem
\bibitem[Hare and Tantillo(2017)Hare, and Tantillo]{hare2017b}
Hare,~S.~R.; Tantillo,~D.~J. Cryptic Post-Transition State Bifurcations that
  Reduce the Efficiency of Lactone-Forming Rh-Carbenoid C--H Insertions.
  \emph{Chem. Sci.} \textbf{2017}, \emph{8}, 1442--1449\relax
\mciteBstWouldAddEndPuncttrue
\mciteSetBstMidEndSepPunct{\mcitedefaultmidpunct}
{\mcitedefaultendpunct}{\mcitedefaultseppunct}\relax
\EndOfBibitem
\bibitem[Hare \latin{et~al.}(2017)Hare, Pemberton, and Tantillo]{hare2017c}
Hare,~S.~R.; Pemberton,~R.~P.; Tantillo,~D.~J. Navigating Past a Fork in the
  Road: Carbocation-$\pi$ Interactions Can Manipulate Dynamic Behavior of
  Reactions Facing Post-Transition-State Bifurcations. \emph{J. Am. Chem. Soc.}
  \textbf{2017}, \emph{139}, 7485--7493\relax
\mciteBstWouldAddEndPuncttrue
\mciteSetBstMidEndSepPunct{\mcitedefaultmidpunct}
{\mcitedefaultendpunct}{\mcitedefaultseppunct}\relax
\EndOfBibitem
\bibitem[Yu \latin{et~al.}(2017)Yu, Chen, Yang, He, Patel, Lam, Liu, and
  Houk]{yu2017}
Yu,~P.; Chen,~T.~Q.; Yang,~Z.; He,~C.~Q.; Patel,~A.; Lam,~Y.-h.; Liu,~C.-Y.;
  Houk,~K. Mechanisms and Origins of Periselectivity of the Ambimodal [6+4]
  Cycloadditions of Tropone to Dimethylfulvene. \emph{J. Am. Chem. Soc.}
  \textbf{2017}, \emph{139}, 8251--8258\relax
\mciteBstWouldAddEndPuncttrue
\mciteSetBstMidEndSepPunct{\mcitedefaultmidpunct}
{\mcitedefaultendpunct}{\mcitedefaultseppunct}\relax
\EndOfBibitem
\bibitem[Hare \latin{et~al.}(2018)Hare, Li, and Tantillo]{hare2018}
Hare,~S.~R.; Li,~A.; Tantillo,~D.~J. Post-Transition State Bifurcations Induce
  Dynamical Detours in Pummerer-Like Reactions. \emph{Chem. Sci.}
  \textbf{2018}, \emph{9}, 8937--8945\relax
\mciteBstWouldAddEndPuncttrue
\mciteSetBstMidEndSepPunct{\mcitedefaultmidpunct}
{\mcitedefaultendpunct}{\mcitedefaultseppunct}\relax
\EndOfBibitem
\bibitem[Xue \latin{et~al.}(2019)Xue, Jamieson, Garcia-Borr{\`a}s, Dong, Yang,
  and Houk]{xue2019}
Xue,~X.-S.; Jamieson,~C.~S.; Garcia-Borr{\`a}s,~M.; Dong,~X.; Yang,~Z.;
  Houk,~K. Ambimodal Trispericyclic Transition State and Dynamic Control of
  Periselectivity. \emph{J. Am. Chem. Soc.} \textbf{2019}, \emph{141},
  1217--1221\relax
\mciteBstWouldAddEndPuncttrue
\mciteSetBstMidEndSepPunct{\mcitedefaultmidpunct}
{\mcitedefaultendpunct}{\mcitedefaultseppunct}\relax
\EndOfBibitem
\bibitem[Liu \latin{et~al.}(2020)Liu, Chen, and Houk]{liu2020}
Liu,~F.; Chen,~Y.; Houk,~K. Huisgen's 1,3-Dipolar Cycloadditions to Fulvenes
  Proceed via Ambimodal [6+4]/[4+2] Transition States. \emph{Angew. Chem., Int.
  Ed.} \textbf{2020}, \emph{59}, 12412--12416\relax
\mciteBstWouldAddEndPuncttrue
\mciteSetBstMidEndSepPunct{\mcitedefaultmidpunct}
{\mcitedefaultendpunct}{\mcitedefaultseppunct}\relax
\EndOfBibitem
\bibitem[Zhang \latin{et~al.}(2021)Zhang, Novak, Jamieson, Xue, Chen, Trauner,
  and Houk]{zhang2021}
Zhang,~H.; Novak,~A.~J.; Jamieson,~C.~S.; Xue,~X.-S.; Chen,~S.; Trauner,~D.;
  Houk,~K. Computational Exploration of the Mechanism of Critical Steps in the
  Biomimetic Synthesis of Preuisolactone A, and Discovery of New Ambimodal
  (5+2)/(4+2) Cycloadditions. \emph{J. Am. Chem. Soc.} \textbf{2021},
  \emph{143}, 6601--6608\relax
\mciteBstWouldAddEndPuncttrue
\mciteSetBstMidEndSepPunct{\mcitedefaultmidpunct}
{\mcitedefaultendpunct}{\mcitedefaultseppunct}\relax
\EndOfBibitem
\bibitem[Fukui(1970)]{fukui1970}
Fukui,~K. Formulation of the Reaction Coordinate. \emph{J. Phys. Chem.}
  \textbf{1970}, \emph{74}, 4161--4163\relax
\mciteBstWouldAddEndPuncttrue
\mciteSetBstMidEndSepPunct{\mcitedefaultmidpunct}
{\mcitedefaultendpunct}{\mcitedefaultseppunct}\relax
\EndOfBibitem
\bibitem[Fukui(1981)]{fukui1981}
Fukui,~K. The Path of Chemical Reactions --- The IRC Approach. \emph{Acc. Chem.
  Res.} \textbf{1981}, \emph{14}, 363--368\relax
\mciteBstWouldAddEndPuncttrue
\mciteSetBstMidEndSepPunct{\mcitedefaultmidpunct}
{\mcitedefaultendpunct}{\mcitedefaultseppunct}\relax
\EndOfBibitem
\bibitem[Zheng and Thiel(2017)Zheng, and Thiel]{zheng2017}
Zheng,~Y.; Thiel,~W. Computational Insights into an Enzyme-Catalyzed [4+2]
  Cycloaddition. \emph{J. Org. Chem.} \textbf{2017}, \emph{82},
  13563--13571\relax
\mciteBstWouldAddEndPuncttrue
\mciteSetBstMidEndSepPunct{\mcitedefaultmidpunct}
{\mcitedefaultendpunct}{\mcitedefaultseppunct}\relax
\EndOfBibitem
\bibitem[Chen \latin{et~al.}(2018)Chen, Zhang, Wu, and Hess~Jr]{chen2018}
Chen,~N.; Zhang,~F.; Wu,~R.; Hess~Jr,~B.~A. Biosynthesis of Spinosyn A: A [4+2]
  or [6+4] Cycloaddition? \emph{ACS Catal.} \textbf{2018}, \emph{8},
  2353--2358\relax
\mciteBstWouldAddEndPuncttrue
\mciteSetBstMidEndSepPunct{\mcitedefaultmidpunct}
{\mcitedefaultendpunct}{\mcitedefaultseppunct}\relax
\EndOfBibitem
\bibitem[Yang \latin{et~al.}(2018)Yang, Yang, Yu, Li, Doubleday, Park, Patel,
  Jeon, Russell, Liu, Russell, and Houk]{yang2018a}
Yang,~Z.; Yang,~S.; Yu,~P.; Li,~Y.; Doubleday,~C.; Park,~J.; Patel,~A.;
  Jeon,~B.-s.; Russell,~W.~K.; Liu,~H.-w. \latin{et~al.}  Influence of Water
  and Enzyme SpnF on the Dynamics and Energetics of the Ambimodal [6+4]/[4+2]
  Cycloaddition. \emph{Proc. Natl. Acad. Sci. U. S. A.} \textbf{2018},
  \emph{115}, E848--E855\relax
\mciteBstWouldAddEndPuncttrue
\mciteSetBstMidEndSepPunct{\mcitedefaultmidpunct}
{\mcitedefaultendpunct}{\mcitedefaultseppunct}\relax
\EndOfBibitem
\bibitem[Yang \latin{et~al.}(2018)Yang, Zou, Yu, Liu, Dong, and
  Houk]{yang2018b}
Yang,~Z.; Zou,~L.; Yu,~Y.; Liu,~F.; Dong,~X.; Houk,~K. Molecular Dynamics of
  the Two-Stage Mechanism of Cyclopentadiene Dimerization: Concerted or
  Stepwise? \emph{Chem. Phys.} \textbf{2018}, \emph{514}, 120--125\relax
\mciteBstWouldAddEndPuncttrue
\mciteSetBstMidEndSepPunct{\mcitedefaultmidpunct}
{\mcitedefaultendpunct}{\mcitedefaultseppunct}\relax
\EndOfBibitem
\bibitem[Yang \latin{et~al.}(2018)Yang, Dong, Yu, Yu, Li, Jamieson, and
  Houk]{yang2018c}
Yang,~Z.; Dong,~X.; Yu,~Y.; Yu,~P.; Li,~Y.; Jamieson,~C.; Houk,~K.
  Relationships Between Product Ratios in Ambimodal Pericyclic Reactions and
  Bond Lengths in Transition Structures. \emph{J. Am. Chem. Soc.}
  \textbf{2018}, \emph{140}, 3061--3067\relax
\mciteBstWouldAddEndPuncttrue
\mciteSetBstMidEndSepPunct{\mcitedefaultmidpunct}
{\mcitedefaultendpunct}{\mcitedefaultseppunct}\relax
\EndOfBibitem
\bibitem[Yang \latin{et~al.}(2019)Yang, Jamieson, Xue, Garcia-Borr{\`a}s,
  Benton, Dong, Liu, and Houk]{yang2019}
Yang,~Z.; Jamieson,~C.~S.; Xue,~X.-S.; Garcia-Borr{\`a}s,~M.; Benton,~T.;
  Dong,~X.; Liu,~F.; Houk,~K. Mechanisms and Dynamics of Reactions Involving
  Entropic Intermediates. \emph{Trends Chem.} \textbf{2019}, \emph{1},
  22--34\relax
\mciteBstWouldAddEndPuncttrue
\mciteSetBstMidEndSepPunct{\mcitedefaultmidpunct}
{\mcitedefaultendpunct}{\mcitedefaultseppunct}\relax
\EndOfBibitem
\bibitem[Zou and Houk(2021)Zou, and Houk]{zou2021}
Zou,~Y.; Houk,~K. Mechanisms and Dynamics of Synthetic and Biosynthetic
  Formation of Delitschiapyrones: Solvent Control of Ambimodal Periselectivity.
  \emph{J. Am. Chem. Soc.} \textbf{2021}, \emph{143}, 11734--11740\relax
\mciteBstWouldAddEndPuncttrue
\mciteSetBstMidEndSepPunct{\mcitedefaultmidpunct}
{\mcitedefaultendpunct}{\mcitedefaultseppunct}\relax
\EndOfBibitem
\bibitem[Jamieson \latin{et~al.}(2021)Jamieson, Sengupta, and
  Houk]{jamieson2021}
Jamieson,~C.~S.; Sengupta,~A.; Houk,~K. Cycloadditions of Cyclopentadiene and
  Cycloheptatriene with Tropones: All \textit{Endo}-[6+4] Cycloadditions are
  Ambimodal. \emph{J. Am. Chem. Soc.} \textbf{2021}, \emph{143},
  3918--3926\relax
\mciteBstWouldAddEndPuncttrue
\mciteSetBstMidEndSepPunct{\mcitedefaultmidpunct}
{\mcitedefaultendpunct}{\mcitedefaultseppunct}\relax
\EndOfBibitem
\bibitem[Wang \latin{et~al.}(2021)Wang, Zhang, Jiang, Wang, Zhou, Chen, Zhang,
  Tan, Ge, Yang, and Liang]{wang2021}
Wang,~X.; Zhang,~C.; Jiang,~Y.; Wang,~W.; Zhou,~Y.; Chen,~Y.; Zhang,~B.;
  Tan,~R.~X.; Ge,~H.~M.; Yang,~Z.~J. \latin{et~al.}  Influence of Water and
  Enzyme on the Post-Transition State Bifurcation of NgnD-Catalyzed Ambimodal
  [6+4]/[4+2] Cycloaddition. \emph{J. Am. Chem. Soc.} \textbf{2021},
  \emph{143}, 21003--21009\relax
\mciteBstWouldAddEndPuncttrue
\mciteSetBstMidEndSepPunct{\mcitedefaultmidpunct}
{\mcitedefaultendpunct}{\mcitedefaultseppunct}\relax
\EndOfBibitem
\bibitem[Qin \latin{et~al.}(2021)Qin, Tremblay, Hong, and Yang]{qin2021}
Qin,~Z.-X.; Tremblay,~M.; Hong,~X.; Yang,~Z.~J. Entropic Path Sampling:
  Computational Protocol to Evaluate Entropic Profile along a Reaction Path.
  \emph{J. Phys. Chem. Lett.} \textbf{2021}, \emph{12}, 10713--10719\relax
\mciteBstWouldAddEndPuncttrue
\mciteSetBstMidEndSepPunct{\mcitedefaultmidpunct}
{\mcitedefaultendpunct}{\mcitedefaultseppunct}\relax
\EndOfBibitem
\bibitem[Ito \latin{et~al.}(2022)Ito, Maeda, and Harabuchi]{ito2022}
Ito,~T.; Maeda,~S.; Harabuchi,~Y. Kinetic Analysis of a Reaction Path Network
  Including Ambimodal Transition States: A Case Study of an Intramolecular
  Diels--Alder Reaction. \emph{J. Chem. Theory Comput.} \textbf{2022},
  \emph{18}, 1663--1671\relax
\mciteBstWouldAddEndPuncttrue
\mciteSetBstMidEndSepPunct{\mcitedefaultmidpunct}
{\mcitedefaultendpunct}{\mcitedefaultseppunct}\relax
\EndOfBibitem
\bibitem[Medvedev \latin{et~al.}(2017)Medvedev, Zeifman, Novikov, Bushmarinov,
  Stroganov, Titov, Chilov, and Svitanko]{medvedev2017}
Medvedev,~M.~G.; Zeifman,~A.~A.; Novikov,~F.~N.; Bushmarinov,~I.~S.;
  Stroganov,~O.~V.; Titov,~I.~Y.; Chilov,~G.~G.; Svitanko,~I.~V. Quantifying
  Possible Routes for SpnF-Catalyzed Formal Diels--Alder Cycloaddition.
  \emph{J. Am. Chem. Soc.} \textbf{2017}, \emph{139}, 3942--3945\relax
\mciteBstWouldAddEndPuncttrue
\mciteSetBstMidEndSepPunct{\mcitedefaultmidpunct}
{\mcitedefaultendpunct}{\mcitedefaultseppunct}\relax
\EndOfBibitem
\bibitem[Barducci \latin{et~al.}(2008)Barducci, Bussi, and
  Parrinello]{barducci2008}
Barducci,~A.; Bussi,~G.; Parrinello,~M. Well-Tempered Metadynamics: A Smoothly
  Converging and Tunable Free-Energy Method. \emph{Phys. Rev. Lett.}
  \textbf{2008}, \emph{100}, 020603\relax
\mciteBstWouldAddEndPuncttrue
\mciteSetBstMidEndSepPunct{\mcitedefaultmidpunct}
{\mcitedefaultendpunct}{\mcitedefaultseppunct}\relax
\EndOfBibitem
\bibitem[Marx and Hutter(2009)Marx, and Hutter]{marx2009}
Marx,~D.; Hutter,~J. \emph{Ab Initio Molecular Dynamics: Basic Theory and
  Advanced Methods}; Cambridge University Press: Cambridge, U.K., 2009\relax
\mciteBstWouldAddEndPuncttrue
\mciteSetBstMidEndSepPunct{\mcitedefaultmidpunct}
{\mcitedefaultendpunct}{\mcitedefaultseppunct}\relax
\EndOfBibitem
\bibitem[Laio and Parrinello(2002)Laio, and Parrinello]{laio2002}
Laio,~A.; Parrinello,~M. Escaping Free-Energy Minima. \emph{Proc. Natl. Acad.
  Sci. U. S. A.} \textbf{2002}, \emph{99}, 12562--12566\relax
\mciteBstWouldAddEndPuncttrue
\mciteSetBstMidEndSepPunct{\mcitedefaultmidpunct}
{\mcitedefaultendpunct}{\mcitedefaultseppunct}\relax
\EndOfBibitem
\bibitem[Piccini \latin{et~al.}(2018)Piccini, Mendels, and
  Parrinello]{piccini2018}
Piccini,~G.; Mendels,~D.; Parrinello,~M. Metadynamics with Discriminants: A
  Tool for Understanding Chemistry. \emph{J. Chem. Theory Comput.}
  \textbf{2018}, \emph{14}, 5040--5044\relax
\mciteBstWouldAddEndPuncttrue
\mciteSetBstMidEndSepPunct{\mcitedefaultmidpunct}
{\mcitedefaultendpunct}{\mcitedefaultseppunct}\relax
\EndOfBibitem
\bibitem[Rizzi \latin{et~al.}(2019)Rizzi, Mendels, Sicilia, and
  Parrinello]{rizzi2019}
Rizzi,~V.; Mendels,~D.; Sicilia,~E.; Parrinello,~M. Blind Search for Complex
  Chemical Pathways Using Harmonic Linear Discriminant Analysis. \emph{J. Chem.
  Theory Comput.} \textbf{2019}, \emph{15}, 4507--4515\relax
\mciteBstWouldAddEndPuncttrue
\mciteSetBstMidEndSepPunct{\mcitedefaultmidpunct}
{\mcitedefaultendpunct}{\mcitedefaultseppunct}\relax
\EndOfBibitem
\bibitem[Schilling \latin{et~al.}(2020)Schilling, Cunha, and
  Luber]{schilling2020}
Schilling,~M.; Cunha,~R.~A.; Luber,~S. Zooming in on the O--O Bond
  Formation—--An Ab Initio Molecular Dynamics Study Applying Enhanced
  Sampling Techniques. \emph{J. Chem. Theory Comput.} \textbf{2020}, \emph{16},
  2436--2449\relax
\mciteBstWouldAddEndPuncttrue
\mciteSetBstMidEndSepPunct{\mcitedefaultmidpunct}
{\mcitedefaultendpunct}{\mcitedefaultseppunct}\relax
\EndOfBibitem
\bibitem[Fu \latin{et~al.}(2021)Fu, Bernasconi, and Liu]{fu2021}
Fu,~Y.; Bernasconi,~L.; Liu,~P. Ab Initio Molecular Dynamics Simulations of the
  S\textsubscript{N}1/S\textsubscript{N}2 Mechanistic Continuum in
  Glycosylation Reactions. \emph{J. Am. Chem. Soc.} \textbf{2021}, \emph{143},
  1577--1589\relax
\mciteBstWouldAddEndPuncttrue
\mciteSetBstMidEndSepPunct{\mcitedefaultmidpunct}
{\mcitedefaultendpunct}{\mcitedefaultseppunct}\relax
\EndOfBibitem
\bibitem[Trizio and Parrinello(2021)Trizio, and Parrinello]{trizio2021}
Trizio,~E.; Parrinello,~M. From Enhanced Sampling to Reaction Profiles.
  \emph{J. Phys. Chem. Lett.} \textbf{2021}, \emph{12}, 8621--8626\relax
\mciteBstWouldAddEndPuncttrue
\mciteSetBstMidEndSepPunct{\mcitedefaultmidpunct}
{\mcitedefaultendpunct}{\mcitedefaultseppunct}\relax
\EndOfBibitem
\bibitem[Raucci \latin{et~al.}(2022)Raucci, Rizzi, and Parrinello]{raucci2022}
Raucci,~U.; Rizzi,~V.; Parrinello,~M. {Discover, Sample, and Refine: Exploring
  Chemistry with Enhanced Sampling Techniques}. \emph{J. Phys. Chem. Lett.}
  \textbf{2022}, \emph{13}, 1424--1430\relax
\mciteBstWouldAddEndPuncttrue
\mciteSetBstMidEndSepPunct{\mcitedefaultmidpunct}
{\mcitedefaultendpunct}{\mcitedefaultseppunct}\relax
\EndOfBibitem
\bibitem[Li \latin{et~al.}(2018)Li, Jia, Pan, Shao, and Mei]{li2018}
Li,~P.; Jia,~X.; Pan,~X.; Shao,~Y.; Mei,~Y. Accelerated Computation of Free
  Energy Profile at Ab Initio Quantum Mechanical/Molecular Mechanics Accuracy
  via a Semi-Empirical Reference Potential. I. Weighted Thermodynamics
  Perturbation. \emph{J. Chem. Theory Comput.} \textbf{2018}, \emph{14},
  5583--5596\relax
\mciteBstWouldAddEndPuncttrue
\mciteSetBstMidEndSepPunct{\mcitedefaultmidpunct}
{\mcitedefaultendpunct}{\mcitedefaultseppunct}\relax
\EndOfBibitem
\bibitem[Piccini and Parrinello(2019)Piccini, and Parrinello]{piccini2019}
Piccini,~G.; Parrinello,~M. Accurate Quantum Chemical Free Energies at
  Affordable Cost. \emph{J. Phys. Chem. Lett.} \textbf{2019}, \emph{10},
  3727--3731\relax
\mciteBstWouldAddEndPuncttrue
\mciteSetBstMidEndSepPunct{\mcitedefaultmidpunct}
{\mcitedefaultendpunct}{\mcitedefaultseppunct}\relax
\EndOfBibitem
\bibitem[Atz \latin{et~al.}(2021)Atz, Grisoni, and Schneider]{atz2021}
Atz,~K.; Grisoni,~F.; Schneider,~G. Geometric Deep Learning on Molecular
  Representations. \emph{Nat. Mach. Intell.} \textbf{2021}, \emph{3},
  1023--1032\relax
\mciteBstWouldAddEndPuncttrue
\mciteSetBstMidEndSepPunct{\mcitedefaultmidpunct}
{\mcitedefaultendpunct}{\mcitedefaultseppunct}\relax
\EndOfBibitem
\bibitem[Kim \latin{et~al.}(2011)Kim, Ruszczycky, Choi, Liu, and Liu]{kim2011}
Kim,~H.~J.; Ruszczycky,~M.~W.; Choi,~S.-h.; Liu,~Y.-n.; Liu,~H.-w. {An
  Enzyme-Catalyzed [4+2] Cycloaddition is a Key Step in the Biosynthesis of
  Spinosyn A}. \emph{Nature} \textbf{2011}, \emph{473}, 109--112\relax
\mciteBstWouldAddEndPuncttrue
\mciteSetBstMidEndSepPunct{\mcitedefaultmidpunct}
{\mcitedefaultendpunct}{\mcitedefaultseppunct}\relax
\EndOfBibitem
\bibitem[Fage \latin{et~al.}(2015)Fage, Isiorho, Liu, Wagner, Liu, and
  Keatinge-Clay]{fage2015}
Fage,~C.~D.; Isiorho,~E.~A.; Liu,~Y.; Wagner,~D.~T.; Liu,~H.-w.;
  Keatinge-Clay,~A.~T. {The Structure of SpnF, a Standalone Enzyme that
  Catalyzes [4+2] Cycloaddition}. \emph{Nat. Chem. Biol.} \textbf{2015},
  \emph{11}, 256--258\relax
\mciteBstWouldAddEndPuncttrue
\mciteSetBstMidEndSepPunct{\mcitedefaultmidpunct}
{\mcitedefaultendpunct}{\mcitedefaultseppunct}\relax
\EndOfBibitem
\bibitem[Valsson \latin{et~al.}(2016)Valsson, Tiwary, and
  Parrinello]{valsson2016}
Valsson,~O.; Tiwary,~P.; Parrinello,~M. Enhancing Important Fluctuations: Rare
  Events and Metadynamics from a Conceptual Viewpoint. \emph{Annu. Rev. Phys.
  Chem.} \textbf{2016}, \emph{67}, 159--184\relax
\mciteBstWouldAddEndPuncttrue
\mciteSetBstMidEndSepPunct{\mcitedefaultmidpunct}
{\mcitedefaultendpunct}{\mcitedefaultseppunct}\relax
\EndOfBibitem
\bibitem[Bussi and Laio(2020)Bussi, and Laio]{bussi2020}
Bussi,~G.; Laio,~A. Using Metadynamics to Explore Complex Free-Energy
  Landscapes. \emph{Nat. Rev. Phys.} \textbf{2020}, \emph{2}, 200--212\relax
\mciteBstWouldAddEndPuncttrue
\mciteSetBstMidEndSepPunct{\mcitedefaultmidpunct}
{\mcitedefaultendpunct}{\mcitedefaultseppunct}\relax
\EndOfBibitem
\bibitem[Iannuzzi \latin{et~al.}(2003)Iannuzzi, Laio, and
  Parrinello]{iannuzzi2003}
Iannuzzi,~M.; Laio,~A.; Parrinello,~M. Efficient Exploration of Reactive
  Potential Energy Surfaces Using Car-Parrinello Molecular Dynamics.
  \emph{Phys. Rev. Lett.} \textbf{2003}, \emph{90}, 238302\relax
\mciteBstWouldAddEndPuncttrue
\mciteSetBstMidEndSepPunct{\mcitedefaultmidpunct}
{\mcitedefaultendpunct}{\mcitedefaultseppunct}\relax
\EndOfBibitem
\bibitem[Pietrucci(2017)]{pietrucci2017}
Pietrucci,~F. Strategies for the exploration of free energy landscapes: Unity
  in diversity and challenges ahead. \emph{Rev. Phys.} \textbf{2017}, \emph{2},
  32--45\relax
\mciteBstWouldAddEndPuncttrue
\mciteSetBstMidEndSepPunct{\mcitedefaultmidpunct}
{\mcitedefaultendpunct}{\mcitedefaultseppunct}\relax
\EndOfBibitem
\bibitem[Branduardi \latin{et~al.}(2012)Branduardi, Bussi, and
  Parrinello]{branduardi2012}
Branduardi,~D.; Bussi,~G.; Parrinello,~M. Metadynamics with Adaptive Gaussians.
  \emph{J. Chem. Theory Comput.} \textbf{2012}, \emph{8}, 2247--2254\relax
\mciteBstWouldAddEndPuncttrue
\mciteSetBstMidEndSepPunct{\mcitedefaultmidpunct}
{\mcitedefaultendpunct}{\mcitedefaultseppunct}\relax
\EndOfBibitem
\bibitem[Zwanzig(1954)]{zwanzig1954}
Zwanzig,~R.~W. High-temperature equation of state by a perturbation method. I.
  Nonpolar gases. \emph{J. Chem. Phys.} \textbf{1954}, \emph{22},
  1420--1426\relax
\mciteBstWouldAddEndPuncttrue
\mciteSetBstMidEndSepPunct{\mcitedefaultmidpunct}
{\mcitedefaultendpunct}{\mcitedefaultseppunct}\relax
\EndOfBibitem
\bibitem[Grimme \latin{et~al.}(2017)Grimme, Bannwarth, and
  Shushkov]{grimme2017}
Grimme,~S.; Bannwarth,~C.; Shushkov,~P. A Robust and Accurate Tight-Binding
  Quantum Chemical Method for Structures, Vibrational Frequencies, and
  Noncovalent Interactions of Large Molecular Systems Parametrized for All
  spd-Block Elements ($Z$ = 1--86). \emph{J. Chem. Theory Comput.}
  \textbf{2017}, \emph{13}, 1989--2009\relax
\mciteBstWouldAddEndPuncttrue
\mciteSetBstMidEndSepPunct{\mcitedefaultmidpunct}
{\mcitedefaultendpunct}{\mcitedefaultseppunct}\relax
\EndOfBibitem
\bibitem[Zhao and Truhlar(2008)Zhao, and Truhlar]{zhao2008}
Zhao,~Y.; Truhlar,~D.~G. The M06 Suite of Density Functionals for Main Group
  Thermochemistry, Thermochemical Kinetics, Noncovalent Interactions, Excited
  States, and Transition Elements: Two New Functionals and Systematic Testing
  of Four M06-Class Functionals and 12 Other Functionals. \emph{Theor. Chem.
  Acc.} \textbf{2008}, \emph{120}, 215--241\relax
\mciteBstWouldAddEndPuncttrue
\mciteSetBstMidEndSepPunct{\mcitedefaultmidpunct}
{\mcitedefaultendpunct}{\mcitedefaultseppunct}\relax
\EndOfBibitem
\bibitem[Ditchfield \latin{et~al.}(1971)Ditchfield, Hehre, and
  Pople]{ditchfield1971}
Ditchfield,~R.; Hehre,~W.~J.; Pople,~J.~A. Self-Consistent Molecular-Orbital
  Methods. IX. An Extended Gaussian-Type Basis for Molecular-Orbital Studies of
  Organic Molecules. \emph{J. Chem. Phys.} \textbf{1971}, \emph{54},
  724--728\relax
\mciteBstWouldAddEndPuncttrue
\mciteSetBstMidEndSepPunct{\mcitedefaultmidpunct}
{\mcitedefaultendpunct}{\mcitedefaultseppunct}\relax
\EndOfBibitem
\bibitem[Hehre \latin{et~al.}(1972)Hehre, Ditchfield, and Pople]{hehre1972}
Hehre,~W.~J.; Ditchfield,~R.; Pople,~J.~A. Self-Consistent Molecular Orbital
  Methods. XII. Further Extensions of Gaussian-Type Basis Sets for Use in
  Molecular Orbital Studies of Organic Molecules. \emph{J. Chem. Phys.}
  \textbf{1972}, \emph{56}, 2257--2261\relax
\mciteBstWouldAddEndPuncttrue
\mciteSetBstMidEndSepPunct{\mcitedefaultmidpunct}
{\mcitedefaultendpunct}{\mcitedefaultseppunct}\relax
\EndOfBibitem
\bibitem[Hariharan and Pople(1973)Hariharan, and Pople]{hariharan1973}
Hariharan,~P.~C.; Pople,~J.~A. The Influence of Polarization Functions on
  Molecular Orbital Hydrogenation Energies. \emph{Theor. Chim. Acta}
  \textbf{1973}, \emph{28}, 213--222\relax
\mciteBstWouldAddEndPuncttrue
\mciteSetBstMidEndSepPunct{\mcitedefaultmidpunct}
{\mcitedefaultendpunct}{\mcitedefaultseppunct}\relax
\EndOfBibitem
\bibitem[Zhang \latin{et~al.}(2020)Zhang, Liu, and Xie]{zhang2020mxm}
Zhang,~S.; Liu,~Y.; Xie,~L. Molecular Mechanics-Driven Graph Neural Network
  with Multiplex Graph for Molecular Structures. \textit{arXiv preprint},
  arXiv:2011.07457, 2020\relax
\mciteBstWouldAddEndPuncttrue
\mciteSetBstMidEndSepPunct{\mcitedefaultmidpunct}
{\mcitedefaultendpunct}{\mcitedefaultseppunct}\relax
\EndOfBibitem
\bibitem[Barone and Cossi(1998)Barone, and Cossi]{barone1998}
Barone,~V.; Cossi,~M. Quantum Calculation of Molecular Energies and Energy
  Gradients in Solution by a Conductor Solvent Model. \emph{J. Phys. Chem. A}
  \textbf{1998}, \emph{102}, 1995--2001\relax
\mciteBstWouldAddEndPuncttrue
\mciteSetBstMidEndSepPunct{\mcitedefaultmidpunct}
{\mcitedefaultendpunct}{\mcitedefaultseppunct}\relax
\EndOfBibitem
\bibitem[Cossi \latin{et~al.}(2003)Cossi, Rega, Scalmani, and
  Barone]{cossi2003}
Cossi,~M.; Rega,~N.; Scalmani,~G.; Barone,~V. Energies, Structures, and
  Electronic Properties of Molecules in Solution With the C-PCM Solvation
  Model. \emph{J. Comput. Chem.} \textbf{2003}, \emph{24}, 669--681\relax
\mciteBstWouldAddEndPuncttrue
\mciteSetBstMidEndSepPunct{\mcitedefaultmidpunct}
{\mcitedefaultendpunct}{\mcitedefaultseppunct}\relax
\EndOfBibitem
\bibitem[Marenich \latin{et~al.}(2012)Marenich, Jerome, Cramer, and
  Truhlar]{marenich2012}
Marenich,~A.~V.; Jerome,~S.~V.; Cramer,~C.~J.; Truhlar,~D.~G. Charge Model 5:
  An Extension of Hirshfeld Population Analysis for the Accurate Description of
  Molecular Interactions in Gaseous and Condensed Phases. \emph{J. Chem. Theory
  Comput.} \textbf{2012}, \emph{8}, 527--541\relax
\mciteBstWouldAddEndPuncttrue
\mciteSetBstMidEndSepPunct{\mcitedefaultmidpunct}
{\mcitedefaultendpunct}{\mcitedefaultseppunct}\relax
\EndOfBibitem
\bibitem[Caruana(1997)]{caruana1997}
Caruana,~R. Multitask Learning. \emph{Mach. Learn.} \textbf{1997}, \emph{28},
  41--75\relax
\mciteBstWouldAddEndPuncttrue
\mciteSetBstMidEndSepPunct{\mcitedefaultmidpunct}
{\mcitedefaultendpunct}{\mcitedefaultseppunct}\relax
\EndOfBibitem
\bibitem[Henkelman and J{\'o}nsson(2000)Henkelman, and
  J{\'o}nsson]{henkelman2000}
Henkelman,~G.; J{\'o}nsson,~H. Improved Tangent Estimate in the Nudged Elastic
  Band Method for Finding Minimum Energy Paths and Saddle Points. \emph{J.
  Chem. Phys.} \textbf{2000}, \emph{113}, 9978--9985\relax
\mciteBstWouldAddEndPuncttrue
\mciteSetBstMidEndSepPunct{\mcitedefaultmidpunct}
{\mcitedefaultendpunct}{\mcitedefaultseppunct}\relax
\EndOfBibitem
\bibitem[Bal \latin{et~al.}(2020)Bal, Fukuhara, Shibuta, and Neyts]{bal2020}
Bal,~K.~M.; Fukuhara,~S.; Shibuta,~Y.; Neyts,~E.~C. Free Energy Barriers from
  Biased Molecular Dynamics Simulations. \emph{J. Chem. Phys.} \textbf{2020},
  \emph{153}, 114118\relax
\mciteBstWouldAddEndPuncttrue
\mciteSetBstMidEndSepPunct{\mcitedefaultmidpunct}
{\mcitedefaultendpunct}{\mcitedefaultseppunct}\relax
\EndOfBibitem
\bibitem[Hartmann and Sch{\"u}tte(2007)Hartmann, and Sch{\"u}tte]{hartmann2007}
Hartmann,~C.; Sch{\"u}tte,~C. Comment on Two Distinct Notions of Free Energy.
  \emph{Phys. D} \textbf{2007}, \emph{228}, 59--63\relax
\mciteBstWouldAddEndPuncttrue
\mciteSetBstMidEndSepPunct{\mcitedefaultmidpunct}
{\mcitedefaultendpunct}{\mcitedefaultseppunct}\relax
\EndOfBibitem
\bibitem[Jeon \latin{et~al.}(2017)Jeon, Ruszczycky, Russell, Lin, Kim, Choi,
  Wang, Liu, Patrick, Russell, \latin{et~al.} others]{jeon2017}
Jeon,~B.-s.; Ruszczycky,~M.~W.; Russell,~W.~K.; Lin,~G.-M.; Kim,~N.;
  Choi,~S.-h.; Wang,~S.-A.; Liu,~Y.-n.; Patrick,~J.~W.; Russell,~D.~H.
  \latin{et~al.}  Investigation of the Mechanism of the SpnF-Catalyzed
  [4+2]-Cycloaddition Reaction in the Biosynthesis of Spinosyn A. \emph{Proc.
  Natl. Acad. Sci. U. S. A.} \textbf{2017}, \emph{114}, 10408--10413\relax
\mciteBstWouldAddEndPuncttrue
\mciteSetBstMidEndSepPunct{\mcitedefaultmidpunct}
{\mcitedefaultendpunct}{\mcitedefaultseppunct}\relax
\EndOfBibitem
\bibitem[Anslyn and Dougherty(2006)Anslyn, and Dougherty]{anslyn2006}
Anslyn,~E.~V.; Dougherty,~D.~A. \emph{Modern Physical Organic Chemistry};
  University Science Books: Sausalito, CA, 2006; pp 407--412\relax
\mciteBstWouldAddEndPuncttrue
\mciteSetBstMidEndSepPunct{\mcitedefaultmidpunct}
{\mcitedefaultendpunct}{\mcitedefaultseppunct}\relax
\EndOfBibitem
\bibitem[Bonati \latin{et~al.}(2020)Bonati, Rizzi, and Parrinello]{bonati2020}
Bonati,~L.; Rizzi,~V.; Parrinello,~M. Data-Driven Collective Variables for
  Enhanced Sampling. \emph{J. Phys. Chem. Lett.} \textbf{2020}, \emph{11},
  2998--3004\relax
\mciteBstWouldAddEndPuncttrue
\mciteSetBstMidEndSepPunct{\mcitedefaultmidpunct}
{\mcitedefaultendpunct}{\mcitedefaultseppunct}\relax
\EndOfBibitem
\bibitem[Bonati \latin{et~al.}(2021)Bonati, Piccini, and
  Parrinello]{bonati2021}
Bonati,~L.; Piccini,~G.; Parrinello,~M. Deep Learning the Slow Modes for Rare
  Events Sampling. \emph{Proc. Natl. Acad. Sci. U. S. A.} \textbf{2021},
  \emph{118}, e2113533118\relax
\mciteBstWouldAddEndPuncttrue
\mciteSetBstMidEndSepPunct{\mcitedefaultmidpunct}
{\mcitedefaultendpunct}{\mcitedefaultseppunct}\relax
\EndOfBibitem
\bibitem[Invernizzi and Parrinello(2020)Invernizzi, and
  Parrinello]{invernizzi2020}
Invernizzi,~M.; Parrinello,~M. Rethinking Metadynamics: From Bias Potentials to
  Probability Distributions. \emph{J. Phys. Chem. Lett.} \textbf{2020},
  \emph{11}, 2731--2736\relax
\mciteBstWouldAddEndPuncttrue
\mciteSetBstMidEndSepPunct{\mcitedefaultmidpunct}
{\mcitedefaultendpunct}{\mcitedefaultseppunct}\relax
\EndOfBibitem
\bibitem[Invernizzi and Parrinello(2022)Invernizzi, and
  Parrinello]{invernizzi2022}
Invernizzi,~M.; Parrinello,~M. Exploration vs Convergence Speed in
  Adaptive-bias Enhanced Sampling. \textit{arXiv preprint}, arXiv:2201.09950,
  2022\relax
\mciteBstWouldAddEndPuncttrue
\mciteSetBstMidEndSepPunct{\mcitedefaultmidpunct}
{\mcitedefaultendpunct}{\mcitedefaultseppunct}\relax
\EndOfBibitem
\bibitem[Tiwary and Parrinello(2013)Tiwary, and Parrinello]{tiwary2013}
Tiwary,~P.; Parrinello,~M. From Metadynamics to Dynamics. \emph{Phys. Rev.
  Lett.} \textbf{2013}, \emph{111}, 230602\relax
\mciteBstWouldAddEndPuncttrue
\mciteSetBstMidEndSepPunct{\mcitedefaultmidpunct}
{\mcitedefaultendpunct}{\mcitedefaultseppunct}\relax
\EndOfBibitem
\bibitem[McCarty \latin{et~al.}(2015)McCarty, Valsson, Tiwary, and
  Parrinello]{mccarty2015}
McCarty,~J.; Valsson,~O.; Tiwary,~P.; Parrinello,~M. Variationally Optimized
  Free-Energy Flooding for Rate Calculation. \emph{Phys. Rev. Lett.}
  \textbf{2015}, \emph{115}, 070601\relax
\mciteBstWouldAddEndPuncttrue
\mciteSetBstMidEndSepPunct{\mcitedefaultmidpunct}
{\mcitedefaultendpunct}{\mcitedefaultseppunct}\relax
\EndOfBibitem
\bibitem[K{\"u}hne \latin{et~al.}(2020)K{\"u}hne, Iannuzzi, Del~Ben, Rybkin,
  Seewald, Stein, Laino, Khaliullin, Sch{\"u}tt, Schiffmann, \latin{et~al.}
  others]{cp2k}
K{\"u}hne,~T.~D.; Iannuzzi,~M.; Del~Ben,~M.; Rybkin,~V.~V.; Seewald,~P.;
  Stein,~F.; Laino,~T.; Khaliullin,~R.~Z.; Sch{\"u}tt,~O.; Schiffmann,~F.
  \latin{et~al.}  CP2K: An Electronic Structure and Molecular Dynamics Software
  Package - Quickstep: Efficient and Accurate Electronic Structure
  Calculations. \emph{J. Chem. Phys.} \textbf{2020}, \emph{152}, 194103\relax
\mciteBstWouldAddEndPuncttrue
\mciteSetBstMidEndSepPunct{\mcitedefaultmidpunct}
{\mcitedefaultendpunct}{\mcitedefaultseppunct}\relax
\EndOfBibitem
\bibitem[Tribello \latin{et~al.}(2014)Tribello, Bonomi, Branduardi, Camilloni,
  and Bussi]{plumed2}
Tribello,~G.~A.; Bonomi,~M.; Branduardi,~D.; Camilloni,~C.; Bussi,~G. PLUMED 2:
  New Feathers for an Old Bird. \emph{Comput. Phys. Commun.} \textbf{2014},
  \emph{185}, 604--613\relax
\mciteBstWouldAddEndPuncttrue
\mciteSetBstMidEndSepPunct{\mcitedefaultmidpunct}
{\mcitedefaultendpunct}{\mcitedefaultseppunct}\relax
\EndOfBibitem
\bibitem[Bussi \latin{et~al.}(2007)Bussi, Donadio, and Parrinello]{bussi2007}
Bussi,~G.; Donadio,~D.; Parrinello,~M. Canonical Sampling Through Velocity
  Rescaling. \emph{J. Chem. Phys.} \textbf{2007}, \emph{126}, 014101\relax
\mciteBstWouldAddEndPuncttrue
\mciteSetBstMidEndSepPunct{\mcitedefaultmidpunct}
{\mcitedefaultendpunct}{\mcitedefaultseppunct}\relax
\EndOfBibitem
\bibitem[Frisch \latin{et~al.}(2016)Frisch, Trucks, Schlegel, Scuseria, Robb,
  Cheeseman, Scalmani, Barone, Petersson, Nakatsuji, Li, Caricato, Marenich,
  Bloino, Janesko, Gomperts, Mennucci, Hratchian, Ortiz, Izmaylov, Sonnenberg,
  Williams-Young, Ding, Lipparini, Egidi, Goings, Peng, Petrone, Henderson,
  Ranasinghe, Zakrzewski, Gao, Rega, Zheng, Liang, Hada, Ehara, Toyota, Fukuda,
  Hasegawa, Ishida, Nakajima, Honda, Kitao, Nakai, Vreven, Throssell,
  Montgomery, Peralta, Ogliaro, Bearpark, Heyd, Brothers, Kudin, Staroverov,
  Keith, Kobayashi, Normand, Raghavachari, Rendell, Burant, Iyengar, Tomasi,
  Cossi, Millam, Klene, Adamo, Cammi, Ochterski, Martin, Morokuma, Farkas,
  Foresman, and Fox]{g16}
Frisch,~M.~J.; Trucks,~G.~W.; Schlegel,~H.~B.; Scuseria,~G.~E.; Robb,~M.~A.;
  Cheeseman,~J.~R.; Scalmani,~G.; Barone,~V.; Petersson,~G.~A.; Nakatsuji,~H.
  \latin{et~al.}  \textit{Gaussian 16}, revision C.01. Gaussian, Inc:
  Wallingford, CT, 2016\relax
\mciteBstWouldAddEndPuncttrue
\mciteSetBstMidEndSepPunct{\mcitedefaultmidpunct}
{\mcitedefaultendpunct}{\mcitedefaultseppunct}\relax
\EndOfBibitem
\bibitem[Klicpera \latin{et~al.}(2020)Klicpera, Giri, Margraf, and
  G{\"u}nnemann]{klicpera2020}
Klicpera,~J.; Giri,~S.; Margraf,~J.~T.; G{\"u}nnemann,~S. Fast and
  Uncertainty-Aware Directional Message Passing for Non-Equilibrium Molecules.
  \textit{arXiv preprint}, arXiv:2011.14115, 2020\relax
\mciteBstWouldAddEndPuncttrue
\mciteSetBstMidEndSepPunct{\mcitedefaultmidpunct}
{\mcitedefaultendpunct}{\mcitedefaultseppunct}\relax
\EndOfBibitem
\bibitem[Kingma and Ba(2014)Kingma, and Ba]{kingma2014}
Kingma,~D.~P.; Ba,~J. Adam: A Method for Stochastic Optimization. \textit{arXiv
  preprint}, arXiv:1412.6980, 2014\relax
\mciteBstWouldAddEndPuncttrue
\mciteSetBstMidEndSepPunct{\mcitedefaultmidpunct}
{\mcitedefaultendpunct}{\mcitedefaultseppunct}\relax
\EndOfBibitem
\end{mcitethebibliography}

\end{document}





\newpage

\begin{figure}[!htbp]
  \includegraphics[width=\linewidth]{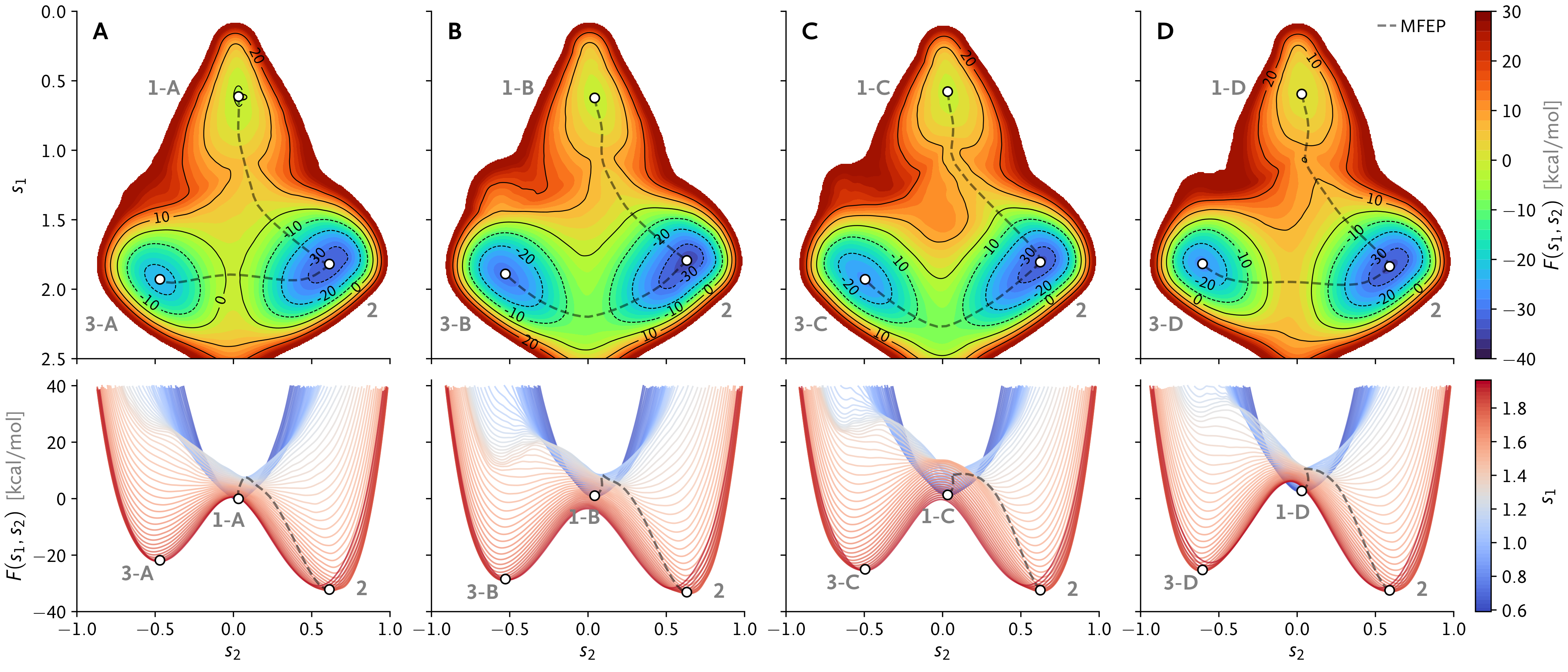}
  \caption{Upper panels: two-dimensional free energy surfaces at the GFN-xTB level, $F(s_1, s_2)$, obtained from the well-tempered metadynamics simulations with different stereochemical restraints (A--D).
  Minimum free energy paths (MFEPs) between the free energy minima are also shown.
  Lower panels: cross-sections of the free energy surfaces along the reaction progress coordinate (CV $s_1$).}
    \label{fig:xtb_fes2d}
\end{figure}

\begin{figure}[!htbp]
  \includegraphics[width=\linewidth]{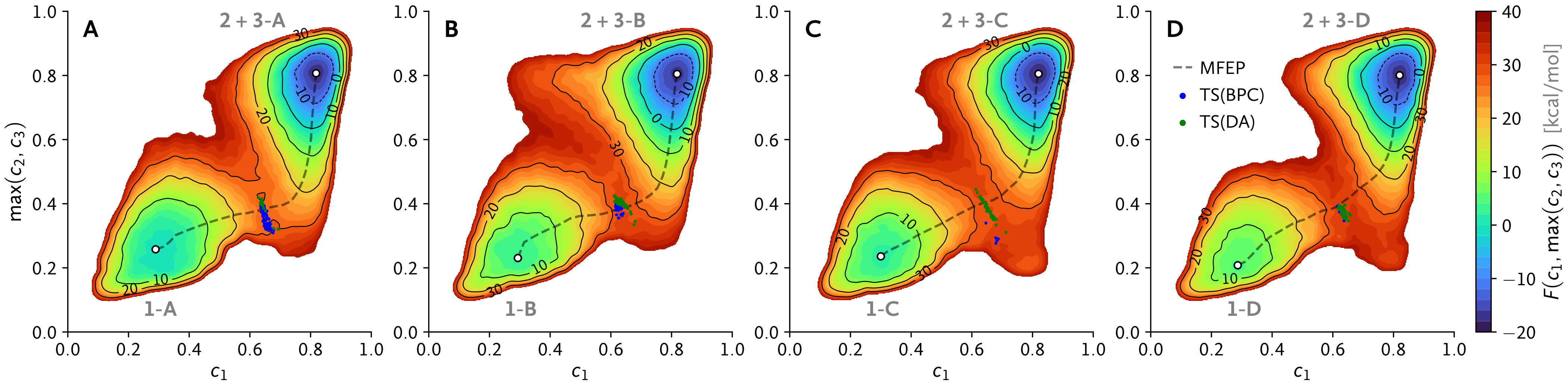}
  \caption{Two-dimensional free energy profiles (More O'Ferrall--Jencks plot) calculated from the well-tempered metadynamics simulations with different stereochemical restraints (A--D).
  The formation of two bonds is represented by the coordination number CVs, $c_1$ and $\max(c_2, c_3)$.
  The minimum free energy path (MFEP) between the free energy minima and transition state structures is also displayed on each plot.}
    \label{fig:mofj_all}
\end{figure}

\begin{figure}[!htbp]
  \includegraphics[width=\linewidth]{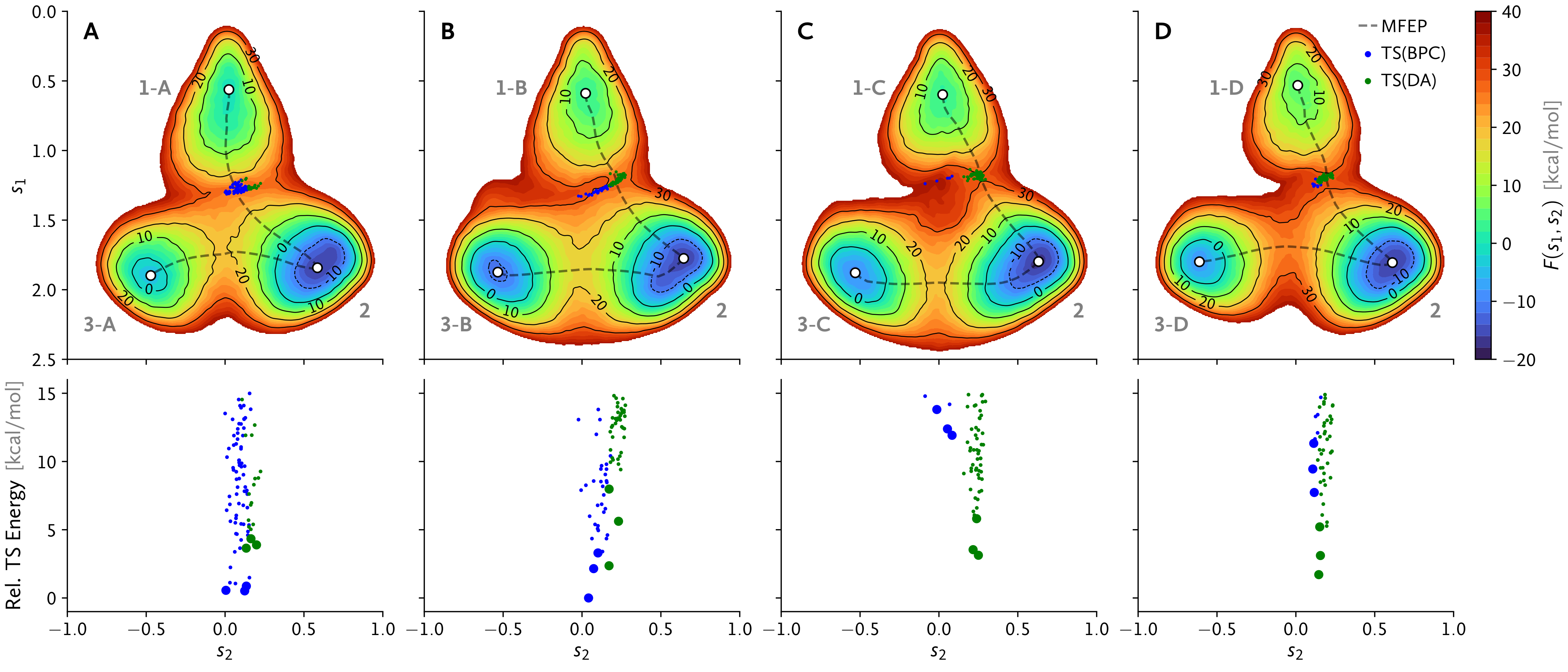}
  \caption{Upper panels: two-dimensional free energy surfaces $F(s_1, s_2)$ obtained from the well-tempered metadynamics simulations with different stereochemical restraints (A--D) (note that these are the same plots shown in Figure~3 in the main text).
  Lower panels: relative energies of the transition states in the different stereochemical pathways obtained at the M06-2X/6-31G(d) level of theory.
  For each stereochemical pathway, the three lowest-energy TSs for each type (BPC and DA) are shown as larger circles.}
  \label{fig:ts_energies}
\end{figure}

\begin{figure}[!htbp]
  \includegraphics[width=\linewidth]{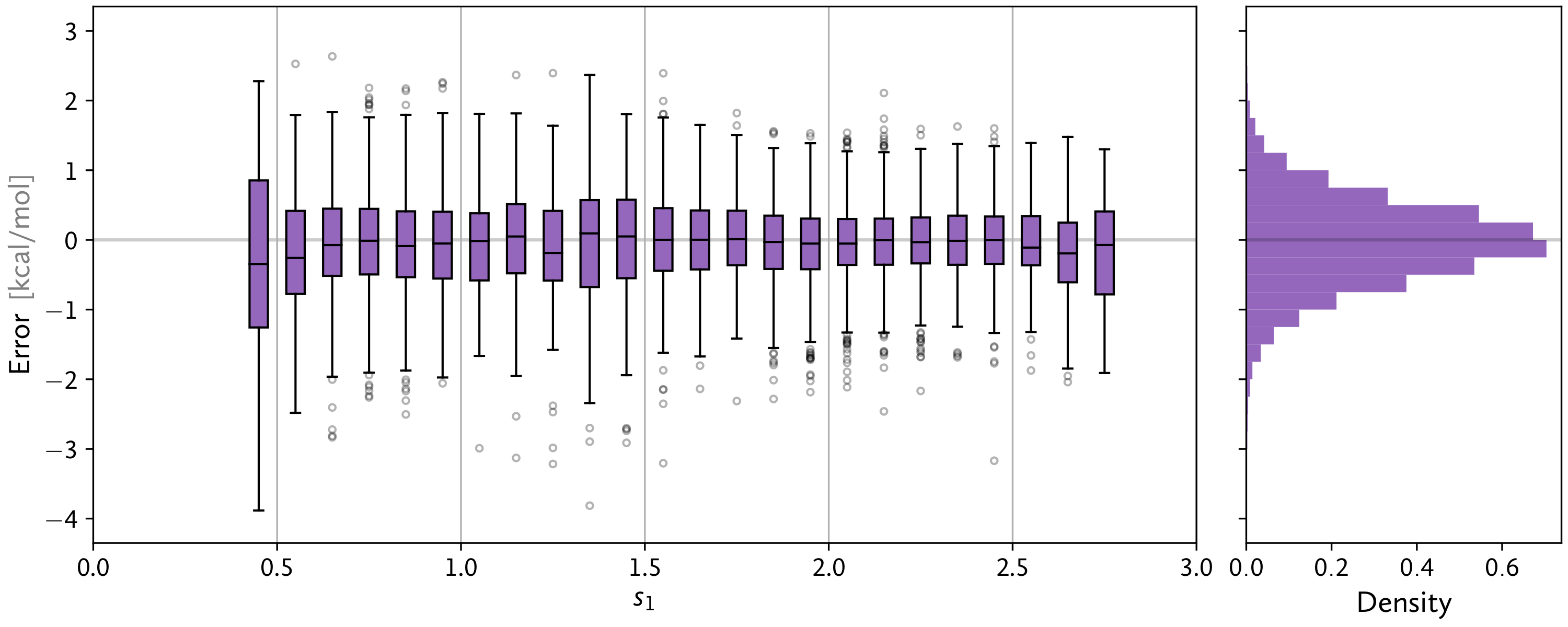}
  \caption{Left panel: error distribution for $E_\text{DFT} - E_\text{xTB}$ over the test set configurations, binned according to CV $s_1$ with a bin size of 0.1.
  The error distribution does not vary significantly with the reaction progress.
  Right panel: histogram of the same error distribution. 
  The absolute prediction error is less than 2 kcal/mol for more than 99\% of the test set configurations.}
    \label{fig:error_hist}
\end{figure}

\begin{table}[!htbp]
  \caption{Free Energies of the Reactant States and Activation Free Energies for the Different Stereochemical Pathways (With Implicit Solvation)}
  \label{table:barrier}
  \begin{tabular}{ccc}
    \hline
    Pathway (X) & $F_{\text{\textbf{1-X}}}$ (kcal/mol) & $\Delta^\ddagger F_\text{X}$ (kcal/mol) \\
    \hline
    A & $3.1 \pm 0.5$ & $28.2 \pm 1.2$ \\
    B & $5.9 \pm 1.0$ & $29.3 \pm 1.4$ \\
    C & $5.8 \pm 1.2$ & $32.0 \pm 2.3$ \\
    D & $7.8 \pm 1.5$ & $33.4 \pm 2.9$ \\
    \hline \\
  \end{tabular}
\end{table}

\begin{table}[!htbp]
  \caption{Comparison of the Mean Absolute Errors of the Target Properties for Different Loss Weights ($w_s$ and $w_c$)}
  \label{table:mae}
  \begin{tabular}{ccc|ccc}
    \hline
    \multirow{3}{*}{$w_s$} & \multirow{3}{*}{$w_c$} & & \multicolumn{3}{c}{Mean Absolute Error (MAE)} \\
    \cline{4-6}
    & & Ensemble & $E_\text{DFT} - E_\text{xTB}$ & $E_\text{DFT/solv} - E_\text{DFT}$ & Partial Charges \\
    & & & (kcal/mol) & (kcal/mol) & ($10^{-4} \, e$) \\
    \hline
    0 & 0 & -- & 0.7819 & -- & -- \\
    0 & 2500 & -- & 0.5764 & -- & 5.09 \\
    1 & 0 & -- & 0.7592 & 0.1902 & -- \\
    1 & 500 & -- & 0.6200 & 0.1268 & 6.69 \\
    1 & 2500 & -- & \textbf{0.5526} & \textbf{0.1172} & 5.02 \\
    1 & 10000 & -- & 0.5662 & 0.1220 & \textbf{4.50} \\
    \hline
    1 & 2500 & $N=5$ & \textbf{0.4855} & \textbf{0.0915} & \textbf{3.72} \\
    \hline
  \end{tabular}
\end{table}


